\pdfoutput=1
\documentclass[aps,prd,reprint,superscriptaddress]{revtex4-1}  

\usepackage{natbib}
\usepackage[pdftex]{graphicx}
\usepackage{color}
\usepackage{amsmath}
\usepackage{subeqnarray}
\usepackage{multirow}

\begin{document}
\title{Subtraction of temperature induced phase noise in the LISA frequency band}
\author{M Nofrarias}  
\email{nofrarias@ice.cat}
\affiliation{Institut de Ci\`encies de l'Espai, (CSIC-IEEC), Facultat de Ci\`encies,
Campus UAB, Torre C-5, 08193 Bellaterra, Spain}
\author{F Gibert}
\affiliation{Institut de Ci\`encies de l'Espai, (CSIC-IEEC), Facultat de Ci\`encies,
Campus UAB, Torre C-5, 08193 Bellaterra, Spain}
\author{N Karnesis}
\affiliation{Institut de Ci\`encies de l'Espai, (CSIC-IEEC), Facultat de Ci\`encies,
Campus UAB, Torre C-5, 08193 Bellaterra, Spain}
\author{A F Garc\'ia } 
\email{Current address: OHB System AG, Universit\"atsallee 27-29, \\ 28359 Bremen, Germany.}
\affiliation{Max-Planck-Institut f\"ur Gravitationsphysik (Albert-Einstein-Institut) and 
Leibniz Universit\"at Hannover, 30167~Hannover, Germany}
\author{M Hewitson}
\affiliation{Max-Planck-Institut f\"ur Gravitationsphysik (Albert-Einstein-Institut) and 
Leibniz Universit\"at Hannover, 30167~Hannover, Germany}
\author{G Heinzel}
\affiliation{Max-Planck-Institut f\"ur Gravitationsphysik (Albert-Einstein-Institut) and 
Leibniz Universit\"at Hannover, 30167~Hannover, Germany}
\author{K Danzmann}
\affiliation{Max-Planck-Institut f\"ur Gravitationsphysik (Albert-Einstein-Institut) and 
Leibniz Universit\"at Hannover, 30167~Hannover, Germany}


\begin{abstract}
Temperature fluctuations are expected to be one of the limiting factors for gravitational
wave detectors in the very low frequency range. Here we report the characterisation of
this noise source in the LISA Pathfinder optical bench and propose a method
to remove its contribution from the data. Our results show that temperature fluctuations
are indeed limiting our measurement below one millihertz, and that their subtraction
leads to a factor 5.6 (15\,dB) reduction in the noise level at the lower end of the LISA
measurement band ($10^{-4}\, \rm Hz$), which increases to 20.2 (26\,dB) at even lower
frequencies, i.e., $1.5\, \times 10^{-5}\, \rm Hz$. The method presented here
can be applied to the subtraction of other noise sources in gravitational wave detectors
in the general situation where multiple sensors are used to characterise the noise source.
\end{abstract}


\maketitle

\section{Introduction}
Temperature fluctuations are expected to be one of the limiting  noise contributions
to gravitational wave interferometers at low frequencies, in the millihertz
band. Currently, ground-based gravitational wave detectors are not limited by
this low frequency contribution because they are already dominated by stronger noise
contributions at low frequencies~\cite{Abadie10}, primarily due to seismic noise
which usually encompasses a variety of effects from human activity, to environmental
microgravity effects. 
This is one of the main driving forces behind the proposal of 
LISA~\cite{Bender00}, a space-borne gravitational
wave detector which aims at observing the gravitational wave sky in
the low frequency region of the spectrum, 
meaning frequencies down to $\rm f \simeq 0.1\,mHz$.
The LISA concept is a constellation of three spacecrafts, each one located at
a vertex of a 5 million kilometers triangle. The constellation follows the Earth in a 
heliocentric orbit, $20^\circ$ behind the Earth, with a $1^\circ$ inclination with respect the ecliptic.
Each spacecraft hosts two test masses in nominally geodesic motion. Inter-spacecraft
laser interferometry is used to measure differential displacements between these test masses, 
which contains the gravitational wave signal. Although ESA and NASA ended 
their collaboration pursuing the implementation of this mission, ESA is currently 
considering a mission which will gather much of the existing LISA heritage, 
the eLISA mission~\cite{Amaro12}, known within ESA as NGO (New Gravitational
wave Observatory). 

For any low frequency space-borne gravitational wave detector, 
the main contributions limiting the instrument performance at low
frequencies are expected to come from spurious accelerations. 
Thermal induced effects are expected to play an important role in 
these contributions. 
Although thermal isolation in the spacecraft 
is expected to account for a $\sim$99\% reduction of the remaining 
solar temperature modulations~\cite{Peabody05}, 
the instrument sensitivity will be directly coupled 
to thermal effects producing forces on the test masses
as is the case of the radiometer
effect, radiation pressure, outgassing~\cite{Carbone07} or, as lately outlined,
brownian gas motion~\cite{Cavalleri09b}. Also, thermal related distortions of the 
optical path or the influence of temperature on other interferometer components, 
such as reference cavities, modulators or photodiodes 
could have an impact on the instrument performance. The ability to prevent or, 
if possible, remove any effect at the low frequency end
of the LISA sensitivity curve is highly desirable since the low frequencies could
contain valuable information about black-hole binaries resulting from mergers
of pre-galactic structures and galaxies~\cite{Bender03}. 

LISA Pathfinder~\cite{Anza05, Armano09} is an ESA mission, with some NASA contributions, 
that will test key technologies required for LISA. In particular, it 
will explore and characterise these thermal related effects around 
1\,mHz.

Ground experiments to test LISA technologies 
and LISA Pathfinder flight hardware 
are the first ones facing this problem and as such,
they are the natural playground to investigate this noise source. 
Below millihertz frequencies, 
daily temperature modulations enter into the measurement
band and are poorly screened since the required isolator for that would need
a high amount mass, assuming a passive insulator~\cite{Lobo06a}.
A different approach is to actively control the thermal fluctuations at low frequencies~\cite{Sanjuan09b},
or in other words, to correct  the slowly varying thermal
drift. However, this shifts the problem to the design of a highly stable
control loop able to measure and remove the low frequency fluctuation with high
precision.

Here we present an alternative to isolation which is to measure the temperature
and subtract its contribution from the main interferometric data stream.
Our method computes the temperature contribution to interferometer data by first
determining the transfer function between both measurements. This is then translated
into a digital filter which is used to compute the noise contribution that
is finally subtracted from the main interferometric measurement. A crucial point
in this scheme is that, in evaluating the phase of the transfer function, we
take into account the group delay between the time of the temperature measurement
and the actual effect in the interferometer so that the subtraction is  performed
coherently. 

The method is general and can be applied to the subtraction of any
noise contribution with a delayed impact on the measurement. To do this, the
variable driving the noise contribution (temperature in our case) must be measured.
As we show below, the error assigned to the subtraction process will be computed
from the coherence between this magnitude and the data once this is subtracted.

This paper is structured as follows. In Sec.~\ref{sec.setup} we describe our experiment setup, 
in Sec.~\ref{sec.charac} we introduce the notation and the basic definitions that we then use to 
characterise our system. In Sec. \ref{sec.analysis} we propose two analysis schemes to 
subtract the temperature noise contributions from the interferometer, which we 
then apply in Sec.~\ref{sec.subt}. We discuss the results and conclude in Sec.~\ref{sec.concl}.

\section{Setup description \label{sec.setup}}
\begin{figure}[t]
\begin{center}
\includegraphics[width=0.49\textwidth]{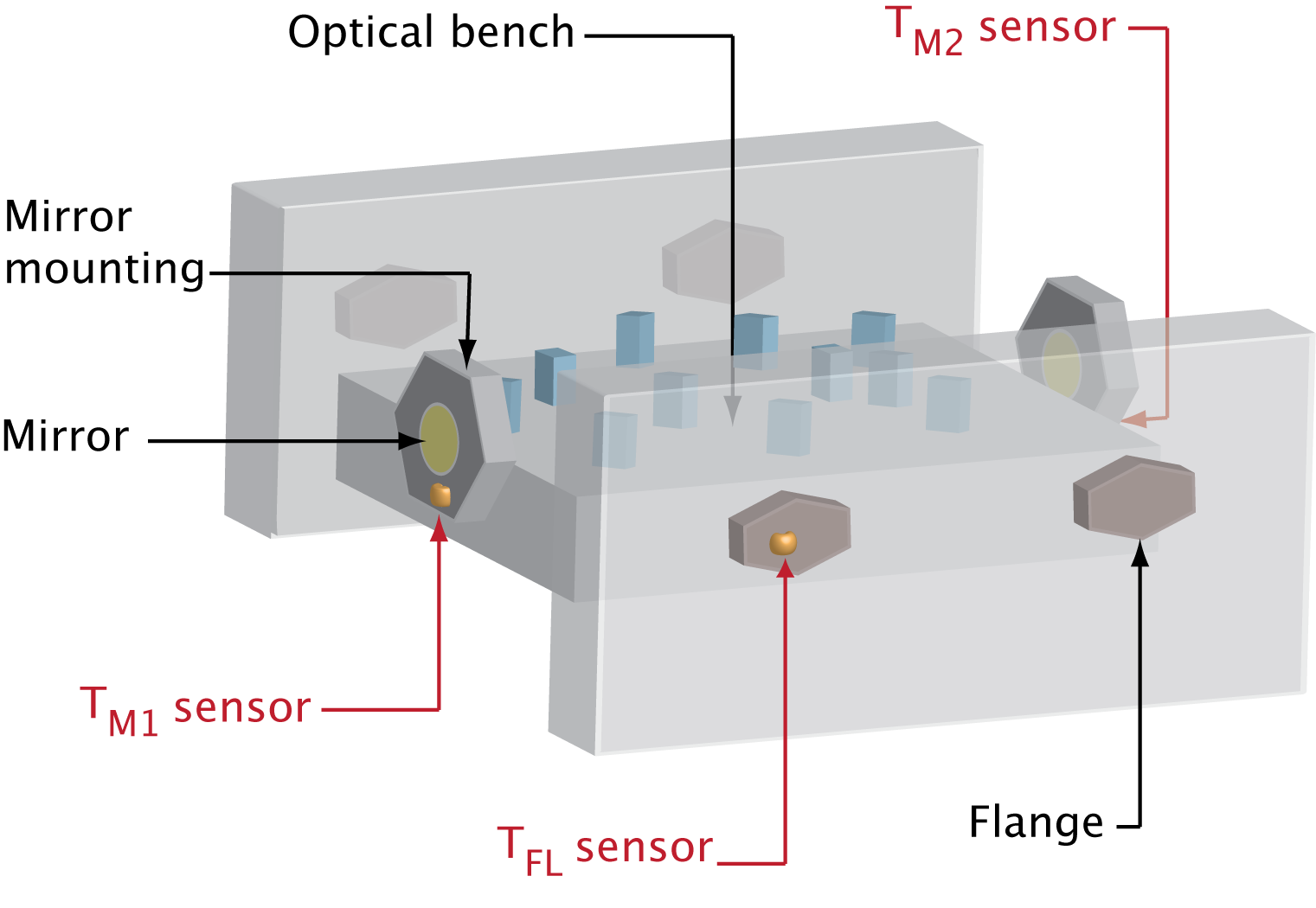}
\caption{ Schematic representation (not to scale) of the location of the temperature
sensors in the LISA Pathfinder optical bench. One sensor was attached 
to each mounting structure supporting the mirrors, a third
sensor was attached to the side slab flange ---a metallic structure 
from which the optical bench will be supported inside the satellite thermal shield---  
and a fourth (not shown) measured the temperature in the laboratory.
\label{fig.scheme}}
\end{center}
\end{figure}
The measurements used in our analysis were taken in the LISA Pathfinder engineering model optical bench~\cite{Heinzel03},
a close replica of the final bench to be flown in the LISA Pathfinder satellite.
In our setup, test masses are substituted by fixed mirrors so only the interferometric
metrology subsystem is tested; none of the required drag-free technology
used in the final mission is part of our setup. 
Although the test-bed has
shown the required performance in the LISA Pathfinder frequency band,
$\rm 1\,mHz \le f \le 1\,Hz$~\cite{Heinzel05, Audley11}, our main aim was to determine
to what extent our measurements were limited by environmental temperature fluctuations and,
also, to evaluate this noise contribution in the LISA band. In order to do that,
we measured the temperature in four different locations of our setup, 
as shown in Fig.~\ref{fig.scheme}.
We initially tested the method using a long data set ($2~\times10^5$\,s),  where the
interferometer was running without active noise suppression control loops.
With the temperature coupling characterised and the method verified, we then
proceeded to subtract the temperature contribution in a shorter, low noise data
segment to characterise the impact of this contribution when the interferometer
is performing according to the mission requirements, reaching 1 $\rm pm/\sqrt{Hz}$
at 0.1\,Hz.

\section{Experiment characterisation \label{sec.charac}}
\begin{figure*}[t]
\centering
$\begin{array}{cccccc}
    &  & \rm T_{Lab} & T_{Fl} & T_{M1} & T_{M2} \\
    & &  &  &  &  \\
																						                                    &
																																 &
\multirow{3}{*}{\includegraphics[width=0.18\textwidth]{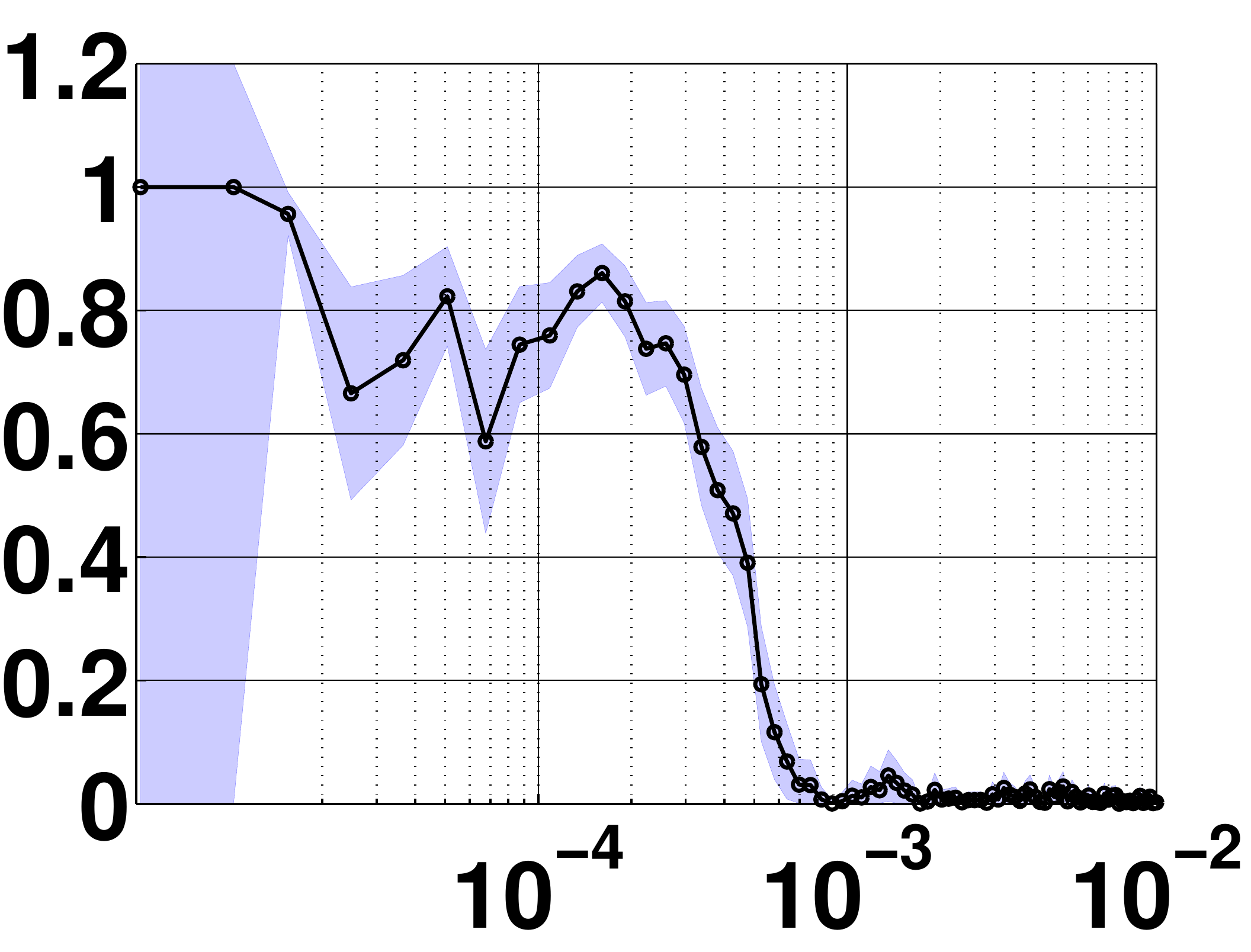}} &
\multirow{3}{*}{\includegraphics[width=0.18\textwidth]{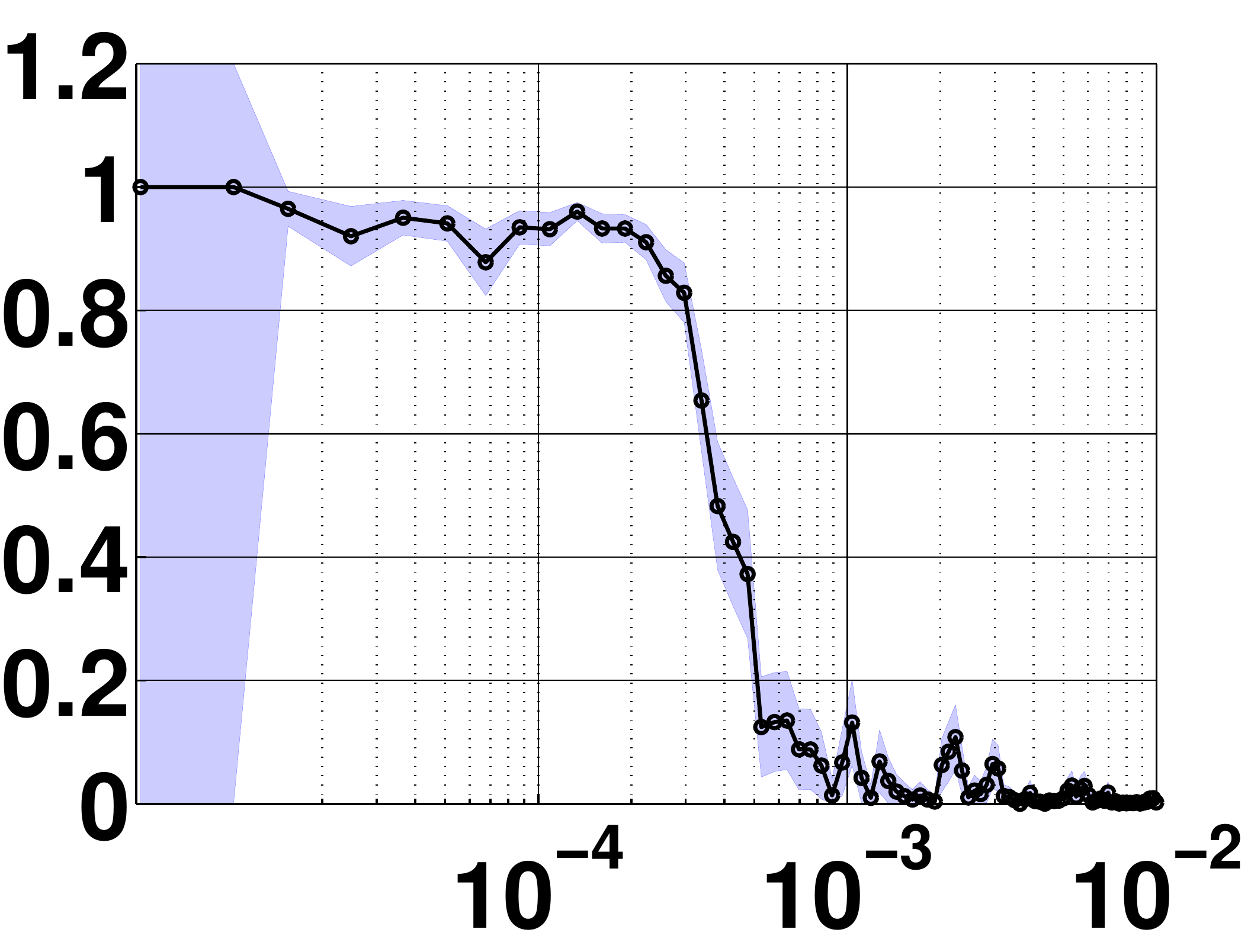}} &
\multirow{3}{*}{\includegraphics[width=0.18\textwidth]{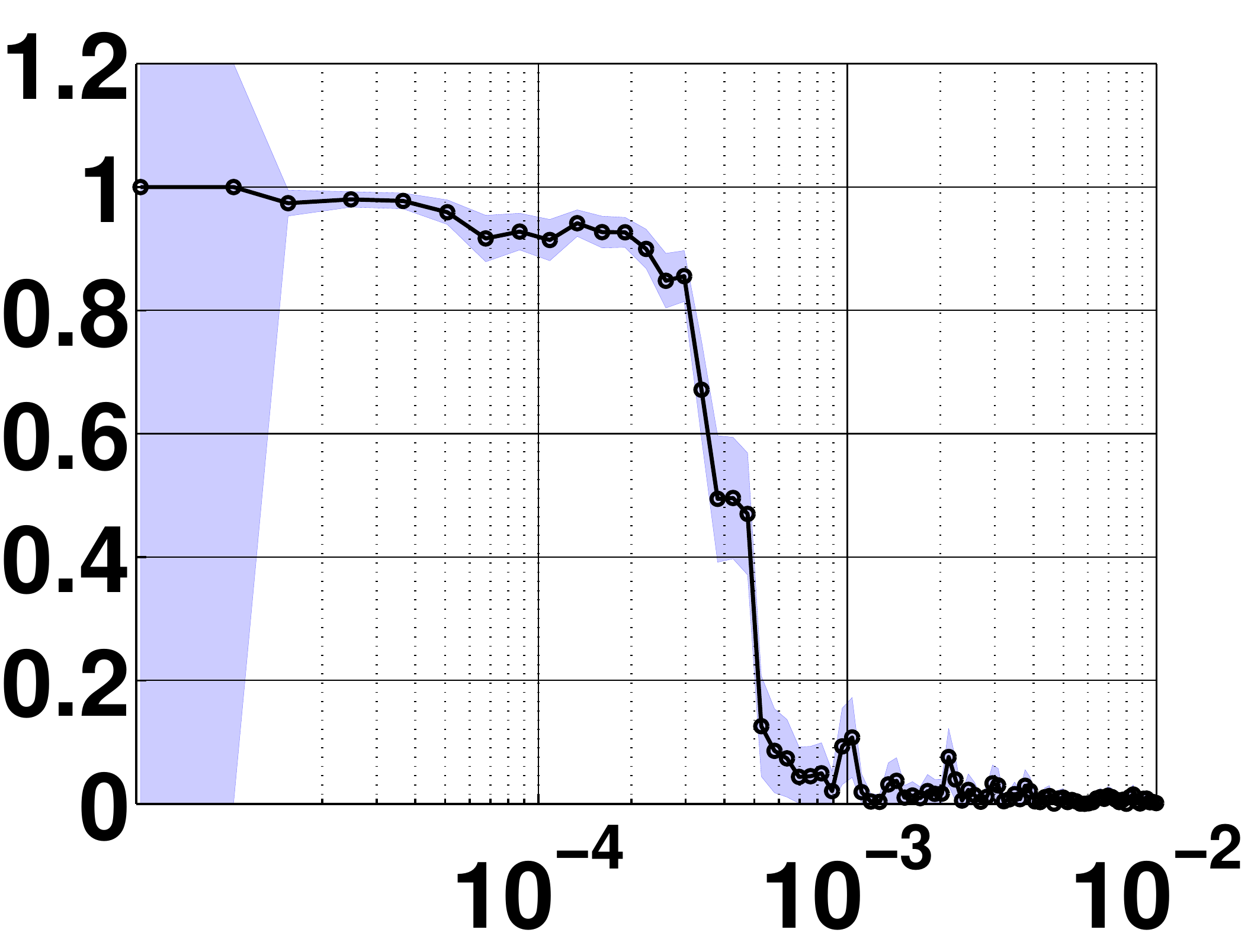}} &
\multirow{3}{*}{\includegraphics[width=0.18\textwidth]{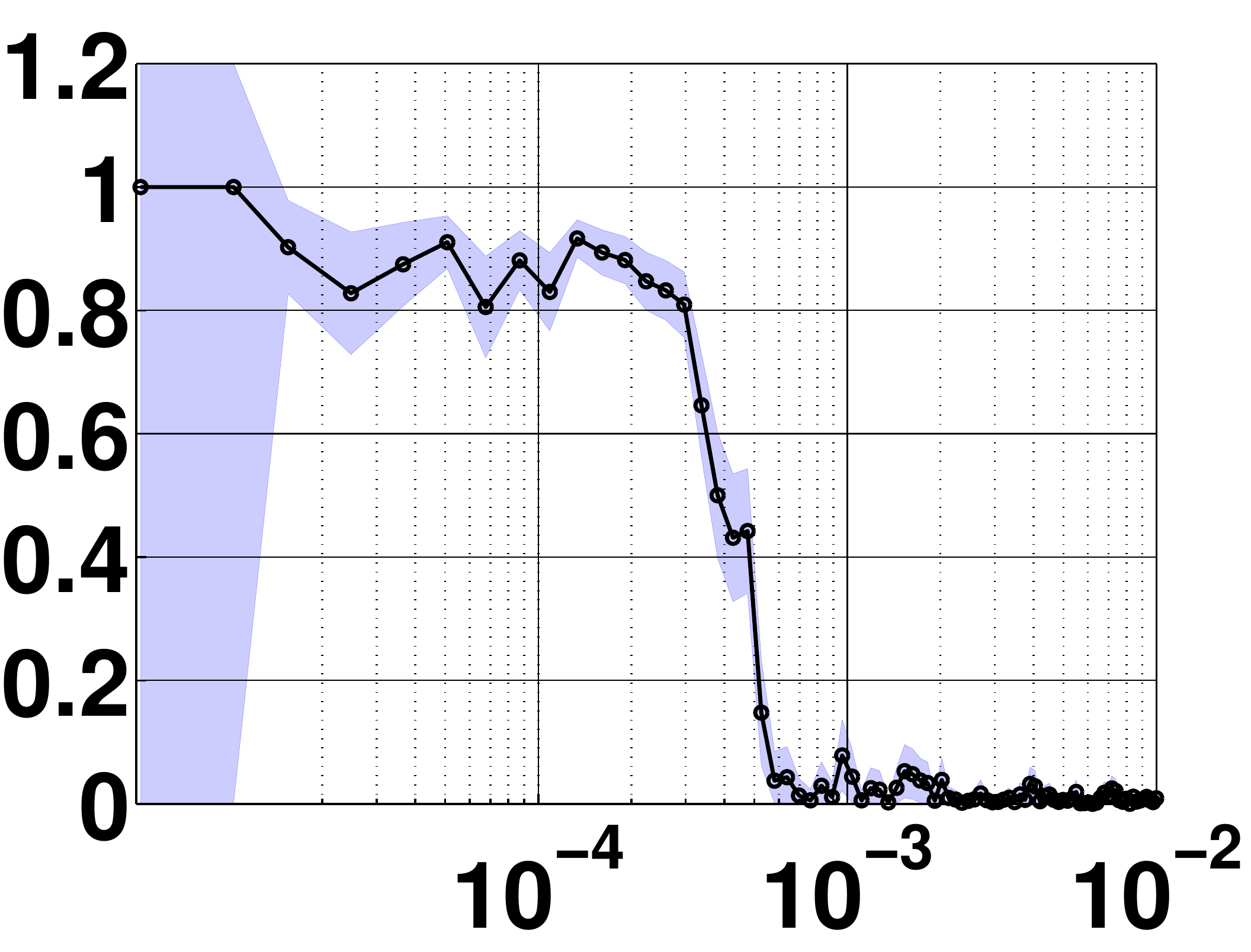}} \\
     & &  &  &  &  \\
  \Phi & &  &  &  &  \\
     & &  &  &  &  \\
     & &  &  &  &  \\
     & &  &  &  &  \\
     & &  &  &  &  \\
    																									   &
																										&
																										&
\multirow{3}{*}{\includegraphics[width=0.17\textwidth]{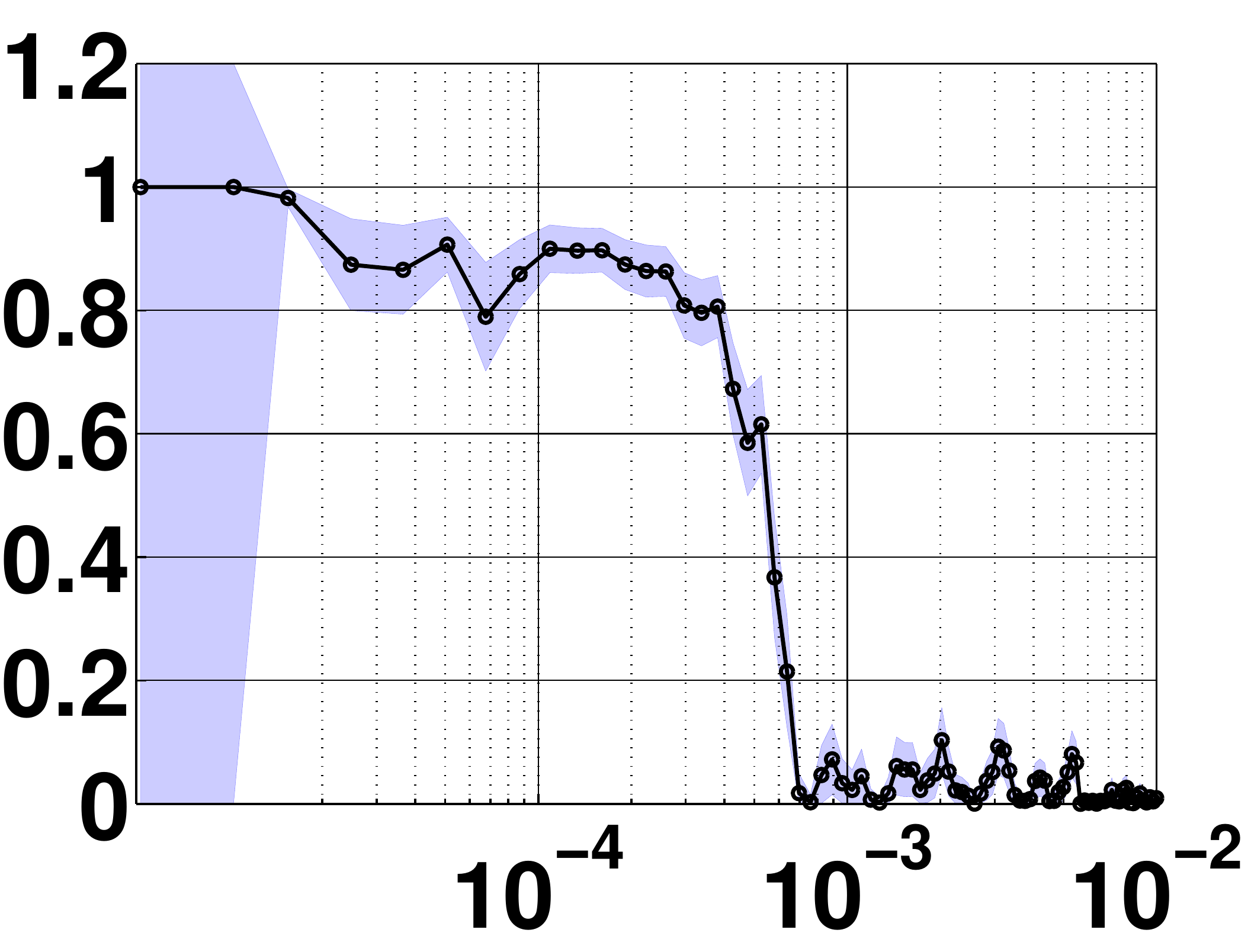}} &
\multirow{3}{*}{\includegraphics[width=0.17\textwidth]{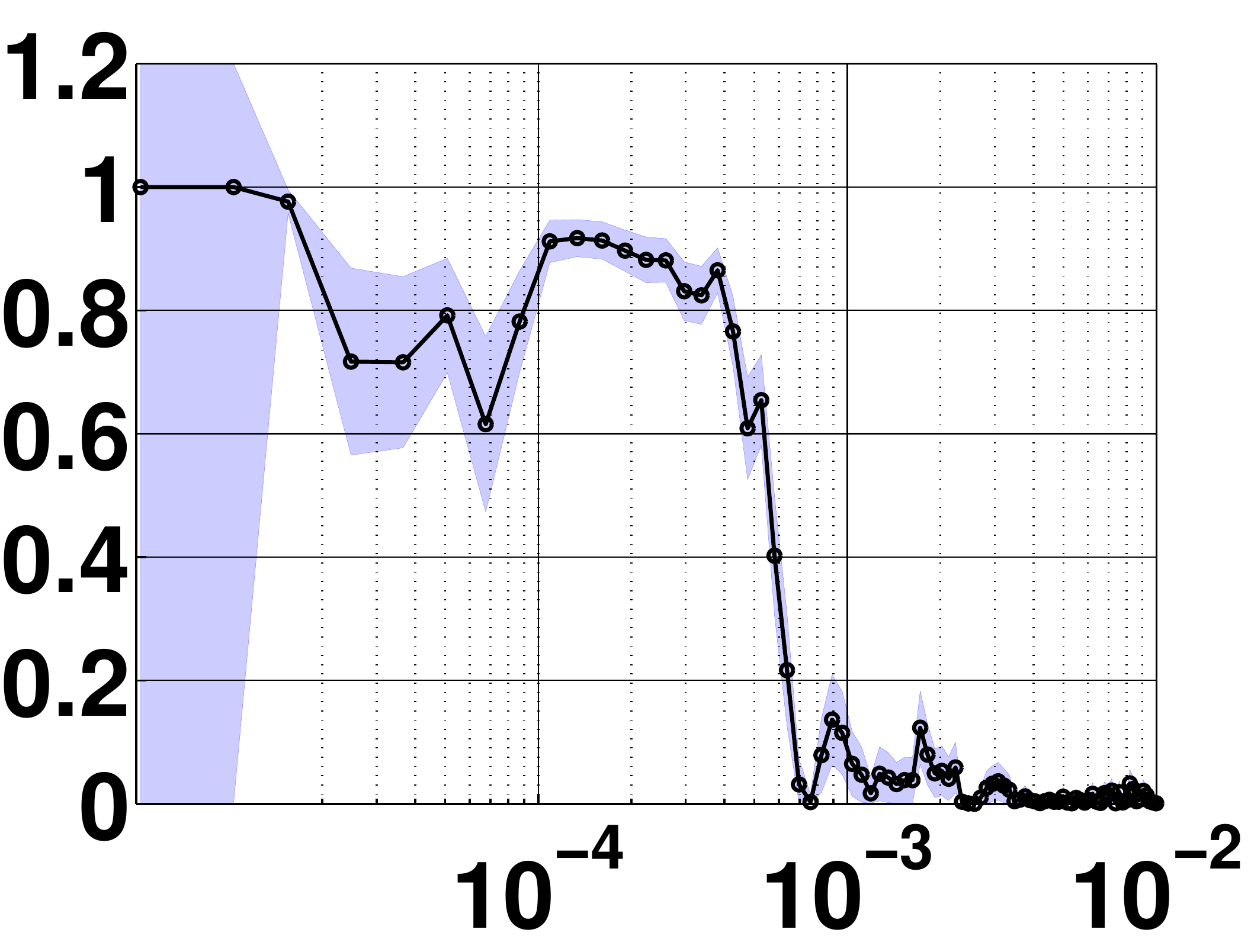}} &
\multirow{3}{*}{\includegraphics[width=0.17\textwidth]{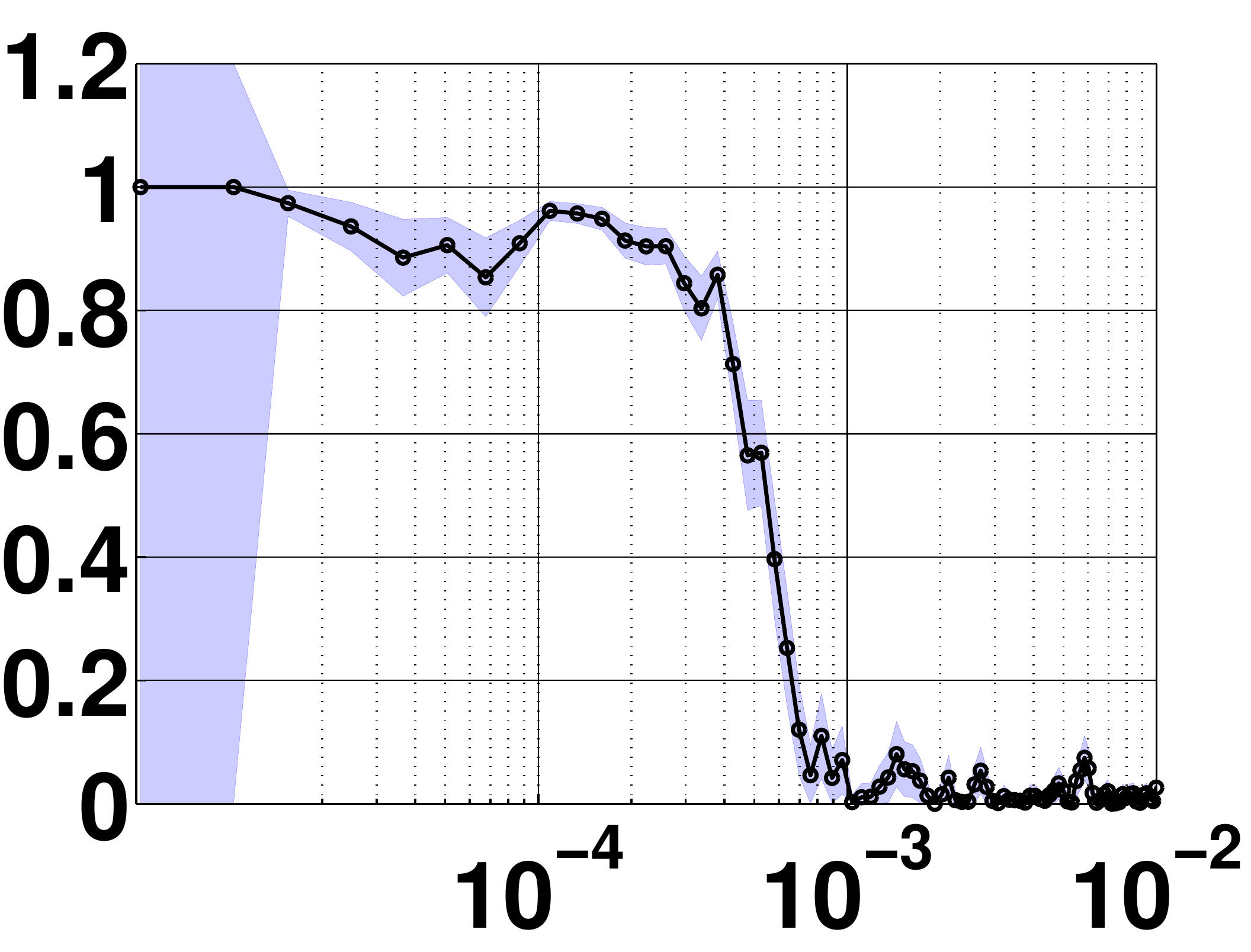}} \\
     & &  &  &  &  \\
\rm T_{Lab}   & &  &  &  &  \\
     & &  &  &  &  \\
     & &  &  &  &  \\
     & &  &  &  &  \\
     & &  &  &  &  \\
																									        &
																											&
																											&
																											&
\multirow{3}{*}{\includegraphics[width=0.17\textwidth]{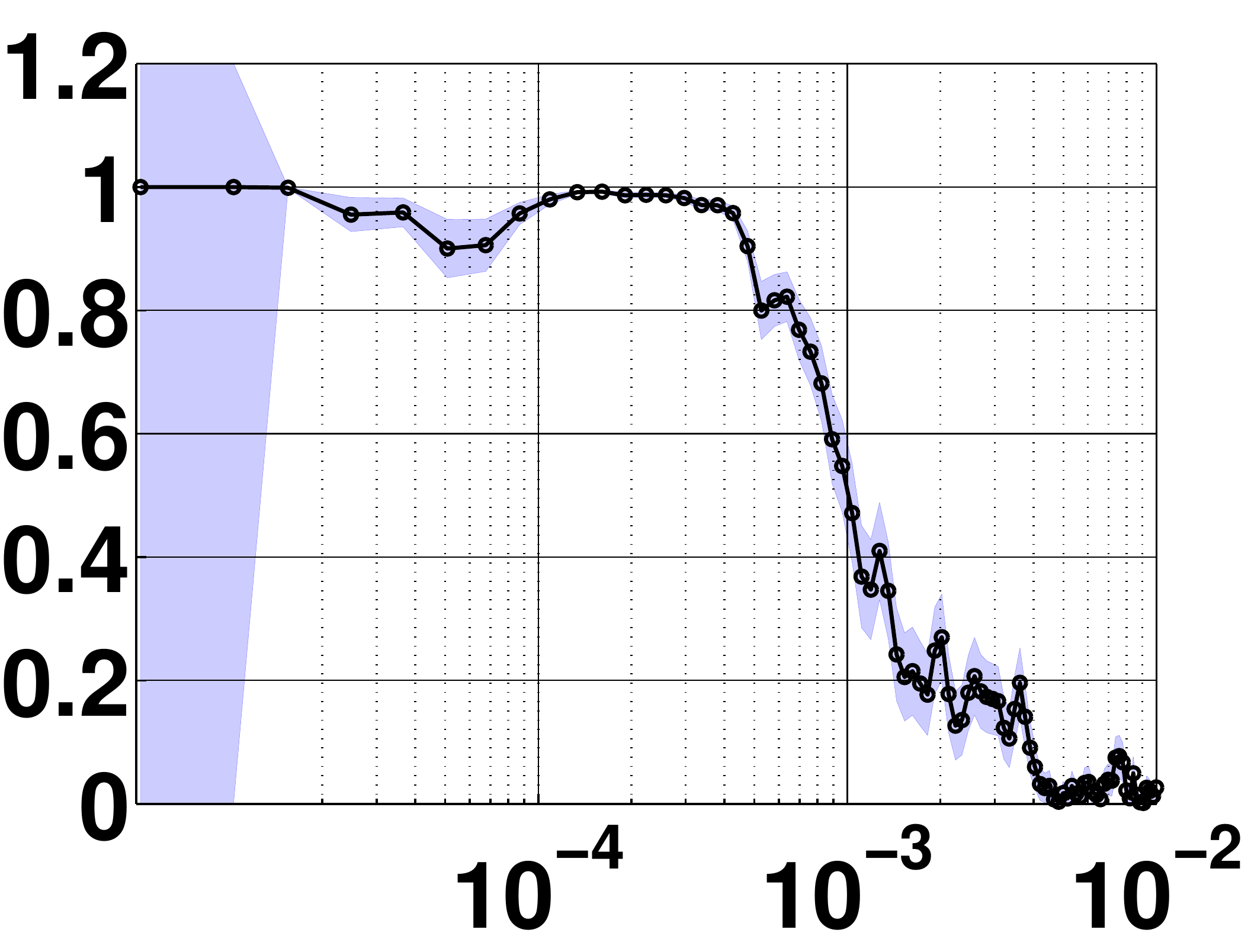}} &
\multirow{3}{*}{\includegraphics[width=0.17\textwidth]{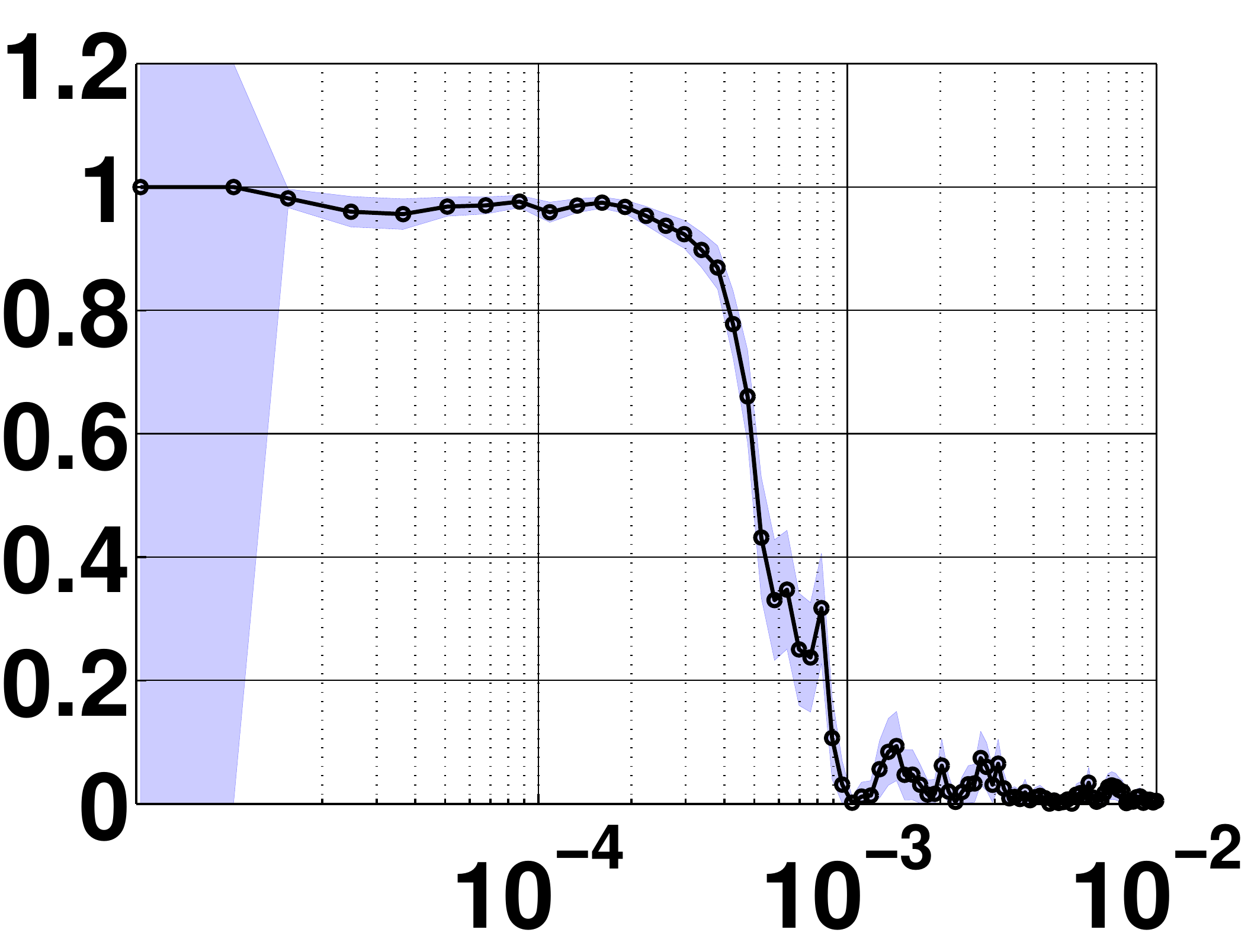}} \\
     & &  &  &  &  \\
\rm T_{Fl}  & &  &  &  &  \\
     & &  &  &  &  \\
     & &  &  &  &  \\
     & &  &  &  &  \\
     & &  &  &  &  \\
																									        &
																											&
																											&
																											&
																											&
\multirow{3}{*}{\includegraphics[width=0.17\textwidth]{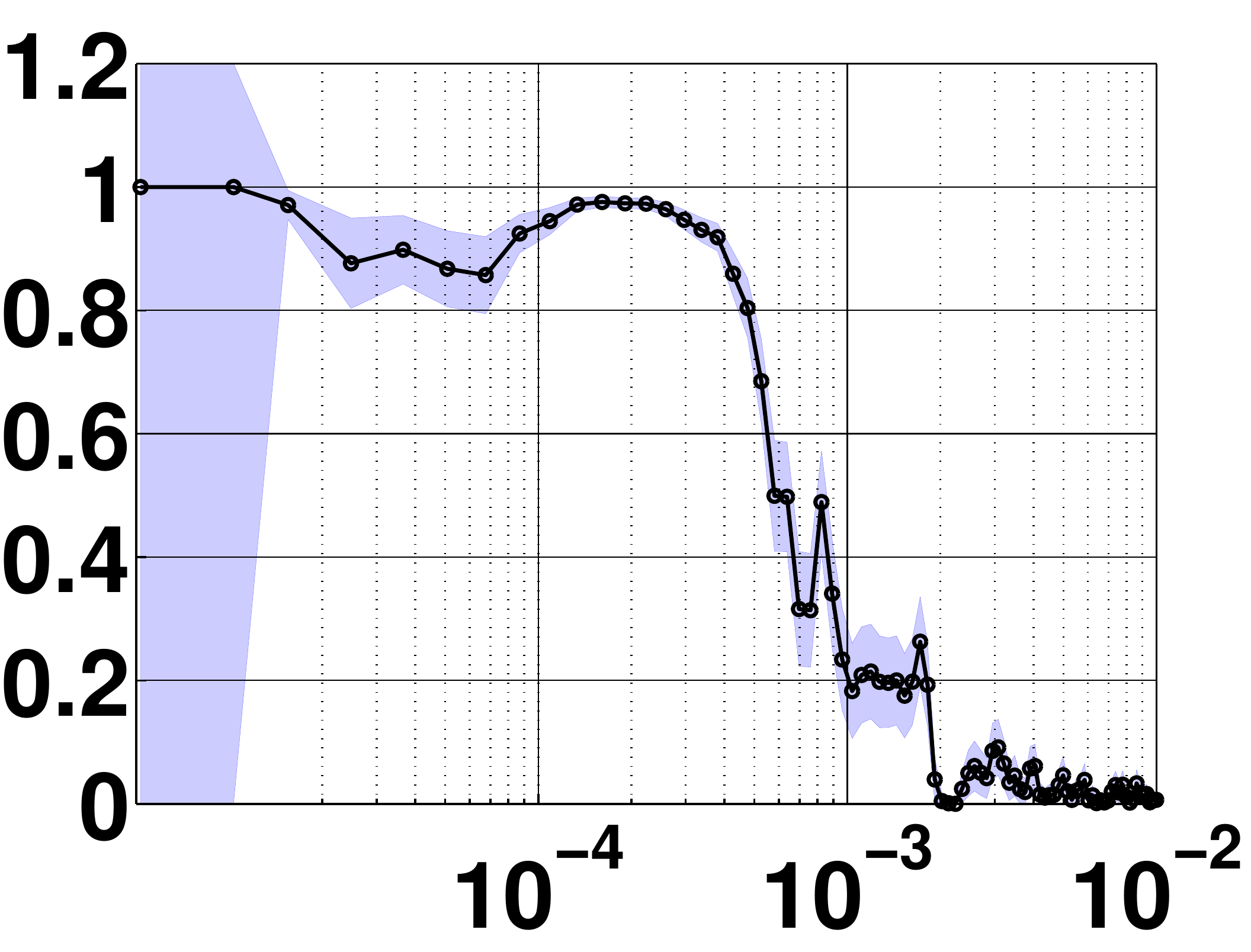}} \\
     & &  &  &  &  \\
\rm T_{M1}  & &  &  &  &  \\
     & &  &  &  &  \\
     & &  &  &  &  \\
     & &  &  &  &  \\
     & &  &  &  &  \\
\end{array}$
\caption{Setup characterisation. We show the matrix of coherence functions between the 
different measurements. In the upper row the coherence between interferometer phase and
temperature sensors, the rest of figures provide the coherence functions between the different 
temperature sensors. Units for the \emph{x}'s values are in hertz.}
\label{fig:coh}
\end{figure*}

\subsection{Signal processing definitions}
We introduce here the basic notation needed to describe 
our experiment. In general, we could describe our experimental data 
as a set of \emph{q} measured temperature time series, 
$T_i(t),\, i=1 \ldots q$, which pass through \emph{q}
systems and combine to produce a single measured 
interferometer output, $\Phi(t)$.
The latter will contain a noise contribution 
which we describe with the random process~$n(t)$.
If we further assume that our $q$ systems are linear systems, this can be
directly translated into the frequency domain as a system of equations
\begin{equation}
\Phi(\omega) = \sum^q_{i=1} H_i(\omega)\,T_i(\omega) + n(\omega),
\label{eq.system}
\end{equation}  
where $\Phi(\omega)$, $T_i(\omega)$ and $n(\omega)$ are the 
Fourier transforms of the inputs, output and noise contribution respectively,
and $H(\omega)$ is the system transfer function. The transfer function is usually estimated from real data 
by~\cite{Bendat}
\begin{equation}
H_{T_i\Phi}(\omega) = \frac{S_{T_i\Phi}(\omega)}{S_{T_iT_i}(\omega)},
\label{eq.transfer}
\end{equation}
where $S_{T_i\Phi}(\omega)$ and $S_{T_iT_i}(\omega)$ are the cross-power spectral 
density and the power spectral density, 
which are defined as the Fourier transform of the correlation functions. 
A second quantity relevant in our case is the coherence function~\cite{Bendat},
\begin{equation}
\gamma^2_{T_i\Phi}(\omega) = \frac{|S_{T_i\Phi}(\omega)|^2}{S_{T_iT_i}(\omega) S_{\Phi\Phi}(\omega)},
\label{eq.coherence}
\end{equation}
which quantifies 
the amount of correlation between two data sets in each frequency bin.
The coherence function turns out to have a particular relevance in our analysis
since it allows us to quantify the error on the estimation of the transfer 
function. 
To bound the uncertainty of our transfer function estimates, 
we compute the error on the transfer
function as~\cite{Bendat}
\begin{equation}
\sigma\left[ | H_{T_i\Phi}| \right]  \simeq  \sigma\left[ \theta ( H_{T_i\Phi}) \right]  \simeq
 \frac{\left[ 1- \gamma^2_{T_i\Phi}(\omega) \right]^{1/2}}{ | \gamma_{T_i\Phi}(\omega)| \, \sqrt{2\,n_d}},\label{eq.errmag}
\end{equation}
where $n_d$ is the number of averages used to compute the transfer function estimates.
The error on the transfer function already
contains the information about the correlation between the temperature and
the interferometer in the form of the coherence function $\gamma^2_{T_i \Phi}(\omega)$.
By doing this, we ensure that our error estimate takes into account the
correlation between temperature and interferometer phase fluctuations.
Notice that the same applies to the correlation between temperatures, where
we will be using the coherence 
$\gamma^2_{T_i T_j}(\omega)$ to evaluate the
interdependence between temperature variations at each sensor.

\subsection{Data preprocessing and characterisation}

Before analysing, the data needed to be resampled onto a common time grid.  Both the
interferometer data (with a sampling frequency of $\rm f_s = 32.47 \, Hz$), and
the temperature data (with $\rm f_s = 1.3 \, Hz$)  were down-sampled to $\rm f_s
= 1\,Hz$.  The temperature sensor data were interpolated to the new time grid with
the new sampling frequency, from 1.3 to 1 \,Hz. In the interferometer case,
the data are first down-sampled by a factor of 10 in order to ease data handling,
and then resampled to the common grid.

To determine the interdependence between temperature and interferometer readout, we compute the coherence functions between them 
$\gamma^2_{T_i \Phi}$. As shown in the upper row of Fig.~\ref{fig:coh}, 
the three sensors attached to the bench show a coherence above 80\% below $f \simeq 0.3$\,mHz. As expected, the sensor measuring temperature variations in the lab has the lowest coherence
with the interferometer, although still a coherence of 80\% can be assigned between interferometer 
and temperature variations in the lab environment. 
While at higher frequencies environmental temperature fluctuations can
be screened by the vacuum tank; in the low frequency region these fluctuations
pass directly through the setup (although delayed, as we show below).

We proceed by determining the interdependence between temperature
measured at different locations. In this case, the coherence function is 
$\gamma^2_{T_i T_j}(\omega)|_{i \ne j}$. 
Results in Fig. \ref{fig:coh} show that the four temperatures are 
strongly correlated ($\gamma^2_{T_{i} T_j}(\omega)
\simeq 1$), in the very low frequency  frequency band ($f < \rm 0.5\,mHz$), and
the correlation drops off rapidly above these frequencies. Hence,
at the low frequencies of interest, the four measurements contain 
mostly the same information regarding coupling between interferometer phase and temperature.

\begin{figure}[t]
\begin{center}
\includegraphics[width=0.5\textwidth]{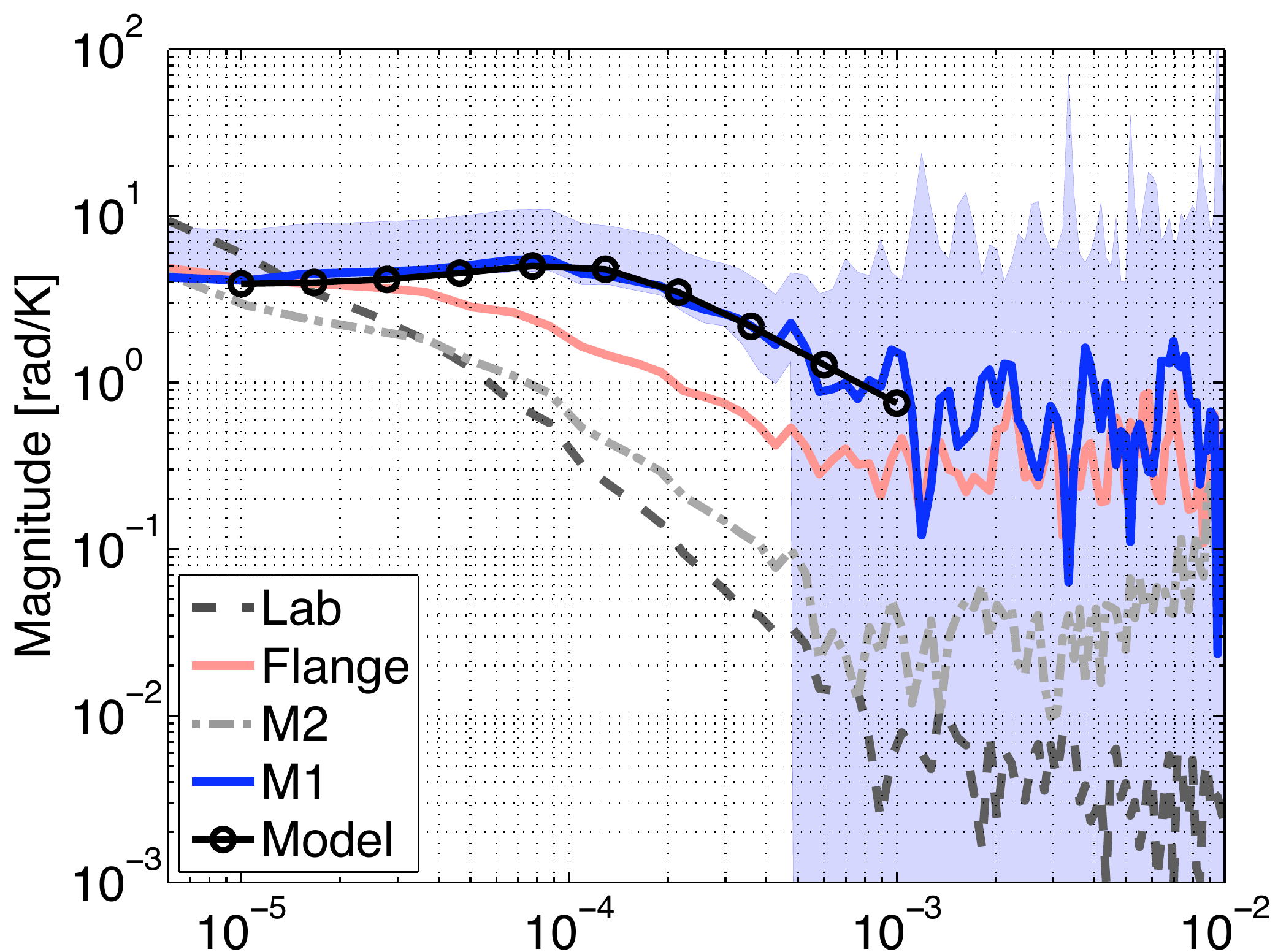}\\[0.1cm]
\includegraphics[width=0.5\textwidth]{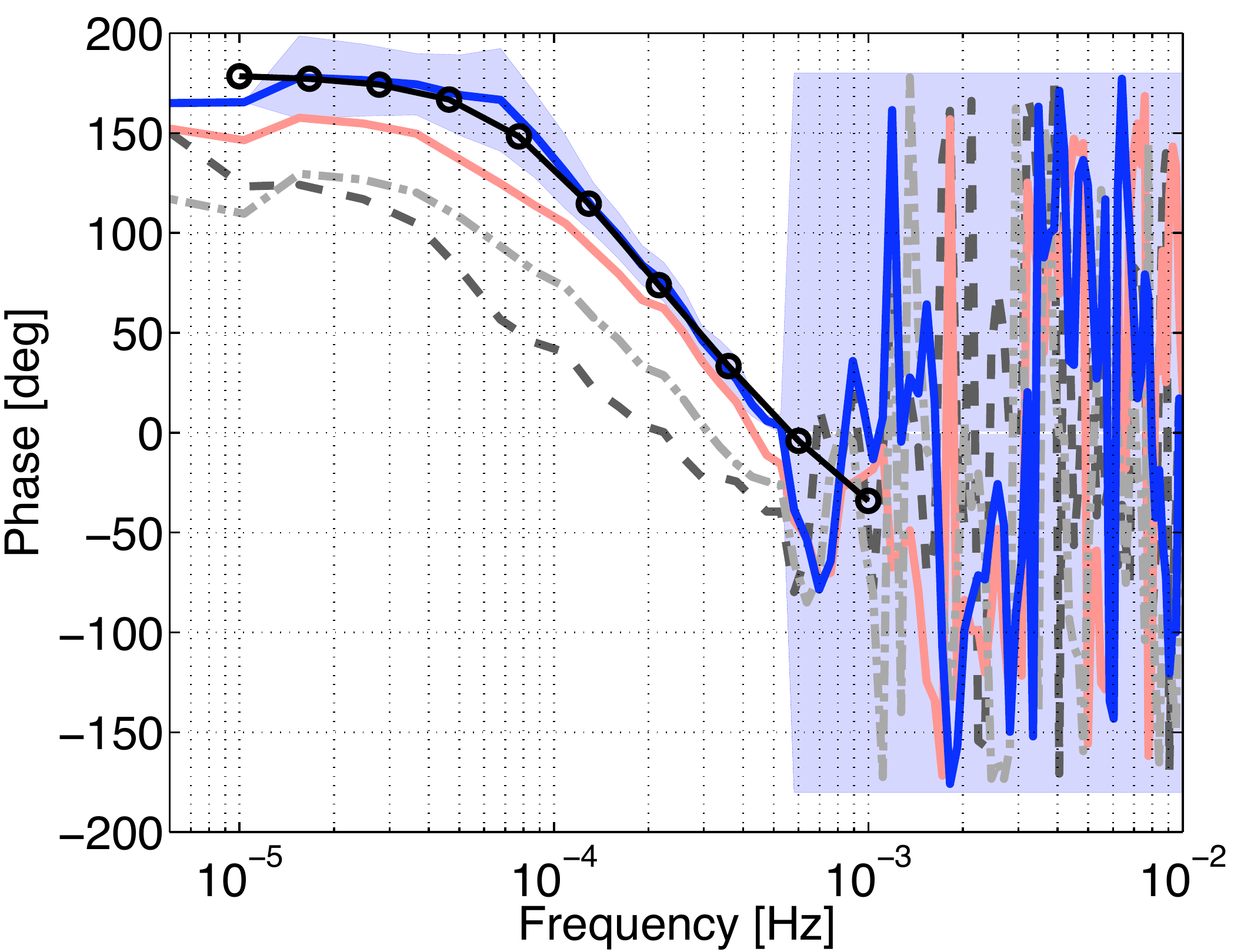}
\caption{Temperature to interferometer transfer functions for the four sensors 
in the setup: Lab environment (Lab), 
optical bench flange (Flange), 
test mass~2 mirror (M2) and 
test mass~1 mirror (M1) ---see  Fig.\ref{fig.scheme} for more details. 
Errors and fitted model are only shown for the transfer function 
that shows the highest contribution, i.e. 
the one coming from the sensor in the first mirror (M1). \label{fig.tf}}
\end{center}
\end{figure}

Even though the four measurements provide similar temperature information at low
frequencies, they do not contribute with the same strength to the interferometric
measurement. To show that, we use the transfer function between the temperature
data and interferometer data ---Eq.~(\ref{eq.transfer}).
In Fig.~\ref{fig.tf} we show these transfer functions for each sensor; the sensor
attached to the first test mass (M1) shows a higher contribution
with respect to the other
---numerical values for
the transfer function are shown in Tab.~\ref{tbl.tf}. 
Also, we computed the group delay of the filter 
by estimating the slope of the phase of the 
transfer function for the M1 sensor in the frequency range 
$\rm 30\,\mu Hz <  f < 0.2\,mHz$.   
In our setup, the response of the interferometer
to temperature fluctuations below the millihertz frequency band is delayed by $1750
\pm 80$\,s with respect to the actual time of temperature measurement.

We note
that, in our analysis, and as shown in Tab. \ref{tbl.tf}, this correlation is valid
for $\rm f \le 0.5\,mHz$. Therefore, the frequency region
where we would expect a noise reduction after subtraction of this contribution would
be below the LISA Pathfinder measuring band, but directly affecting a LISA-like measurement.

\begin{table}[b]
\begin{center}
\caption{Numerical values for the transfer function corresponding to the sensor in the
first mirror mount (M1) and associated errors. 
Last column shows the coherence function. \label{tbl.tf}}
\begin{tabular}{lll} 
\hline\noalign{\smallskip}
Frequency [Hz] \hspace*{0.5cm} &  $|\hat H_{\varphi T}|$ [rad/K] \hspace*{0.5cm} & $\hat \gamma_{\varphi T}$  \\
\noalign{\smallskip}\hline\noalign{\smallskip}
 $1.6 \times 10^{-5}$    & $4.5 \pm 0.5$   & 0.97 \\
 $5.1 \times 10^{-5}$    & $5.1  \pm 0.6$  & 0.96 \\
 $1.1 \times 10^{-4}$    & $4.5 \pm 0.6$   & 0.91  \\
 $2.6 \times 10^{-4}$    & $2.7 \pm 0.5$   & 0.85\\
 $5.3 \times 10^{-4}$    & $1.6 \pm 2.8$   & 0.12 \\
 $1.0 \times 10^{-3}$    &  $1.4 \pm 2.6$  & 0.10  \\
 $2.5 \times 10^{-3}$    & $0.5 \pm 6.5$ & 0.01\\
\noalign{\smallskip}\hline
\end{tabular}
\end{center}
\end{table}

%

\section{Spectral analysis \label{sec.analysis}}

Once we have characterised our system, our next step is to disentangle the
temperature contribution from the interferometer measurement. We will proceed in
two different ways: firstly, by solving the linear system and obtaining a set of
optimal transfer functions and secondly, by applying a conditioned scheme, which
proceeds by solving the linear system for each individual contribution. We
describe both in the following section. 

\subsection{Optimal spectral analysis \label{sec.optimal}}

In order to obtain the system of linear equations that will describe our problem we start
with Eq.~(\ref{eq.system}),
\begin{equation}
\Phi(\omega) = \sum^q_{i=1} H_i(\omega)\,T_i(\omega) + n(\omega),
\end{equation}  
where, if we multiply $\Phi(\omega)$ by its complex conjugate $\Phi(\omega)^*$, 
and taking expectation values we obtain 
\begin{equation}
S_{nn} = S_{\Phi\Phi} - \sum_{i=1}^q H_i\,S_{\Phi T_i} - \sum_{j=1}^q H^*_j\,S_{T_j \Phi} + \sum_{i=j}^q\sum_{j=1}^q H^*_j\,H_i\,S_{T_j 
T_i}. \label{eq.Snn}
\end{equation}
The optimal solution for this system of equations which minimise $S_{nn}$ 
over all possible choices of $H_i$ is found by setting 
\begin{equation}
\frac{\partial S_{nn}}{\partial H_j} = 0 , \quad \frac{\partial S_{nn}}{\partial H^*_j} = 
0,
\end{equation}
which leads to~\cite{Bendat}
\begin{equation}
S_{T_j \Phi} = \sum_{i=1}^q H_i\,S_{T_j T_i} , \quad j = 1, 2, \ldots q,
\label{eq.Hoptimal}
\end{equation}
which can be solved for $H_i$, since $S_{\Phi T_i}$  and $S_{T_i T_j}$ are known. In our particular case  this turns into a system of q = 4 equations, which leads to analytical solutions for $H_i,\; i = 1, \ldots q$. Deriving the expressions
for $H_i$, which we do not reproduce here, requires us to compute 24 terms for each
transfer function, each term containing a product of 4 spectral densities. In
the following section we introduce a sequential method which allows us to handle
simpler equations, leading to equivalent results.

\subsection{Conditioned spectral analysis}
\begin{figure}[t]
\begin{center}
\includegraphics[width=0.5\textwidth]{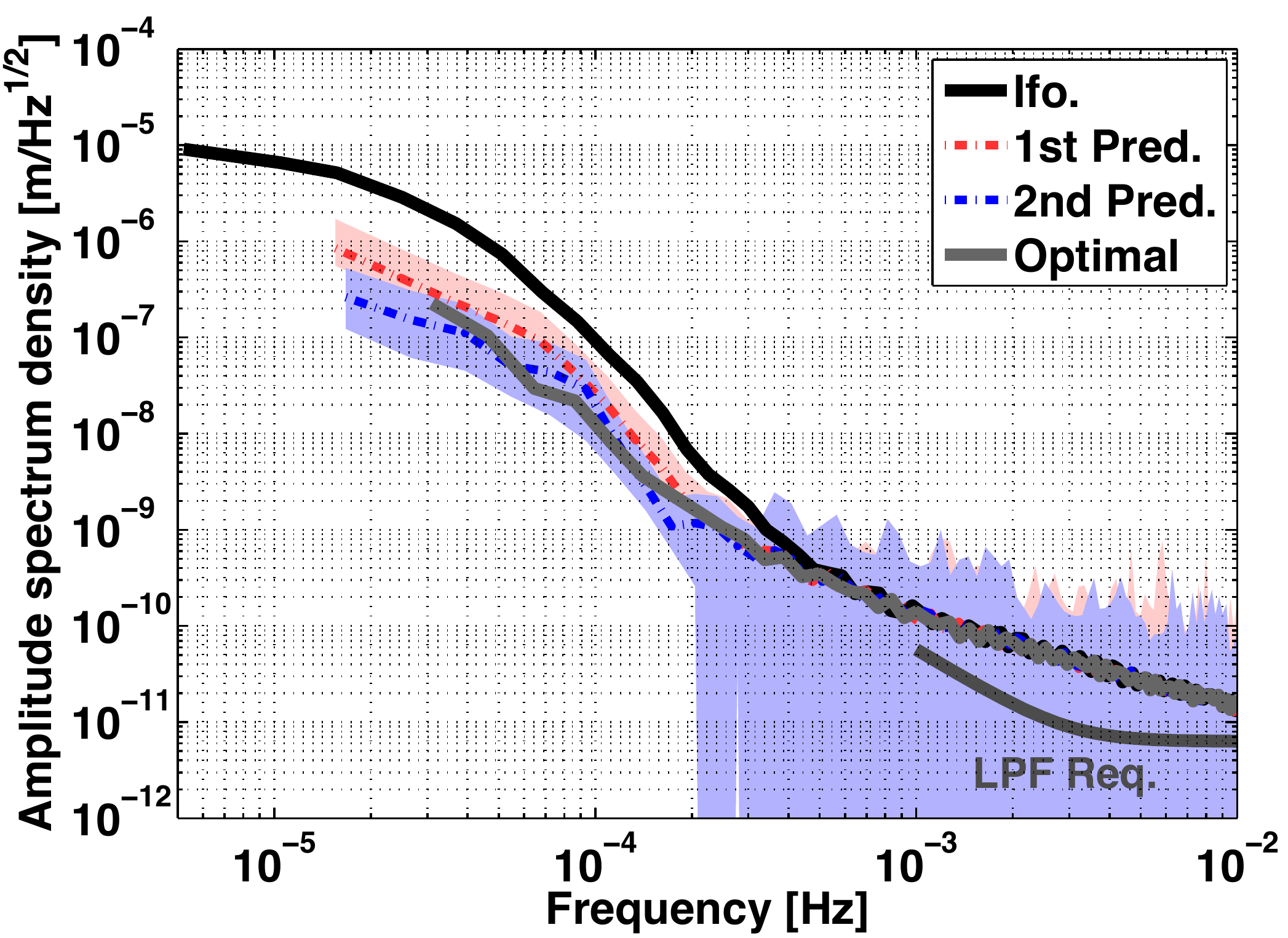}
\caption{Comparison between optimal and conditioned analysis. 
The dashed lines (labelled as 1st Pred. and 2nd Pred.) show the expected noise level after subtraction
of the two sensors with higher contribution, according to the conditioned analysis scheme.
The solid grey (Optimal) line is the expected noise level after an optimal subtraction of all sensors. 
We show in the same figure the 
LISA Pathfinder requirement as a reference. \label{fig.cond2opt}}
\label{fig.subs}
\end{center}
\end{figure} 

\begin{figure*}[t!]
\begin{center}
\includegraphics[width=0.55\textwidth]{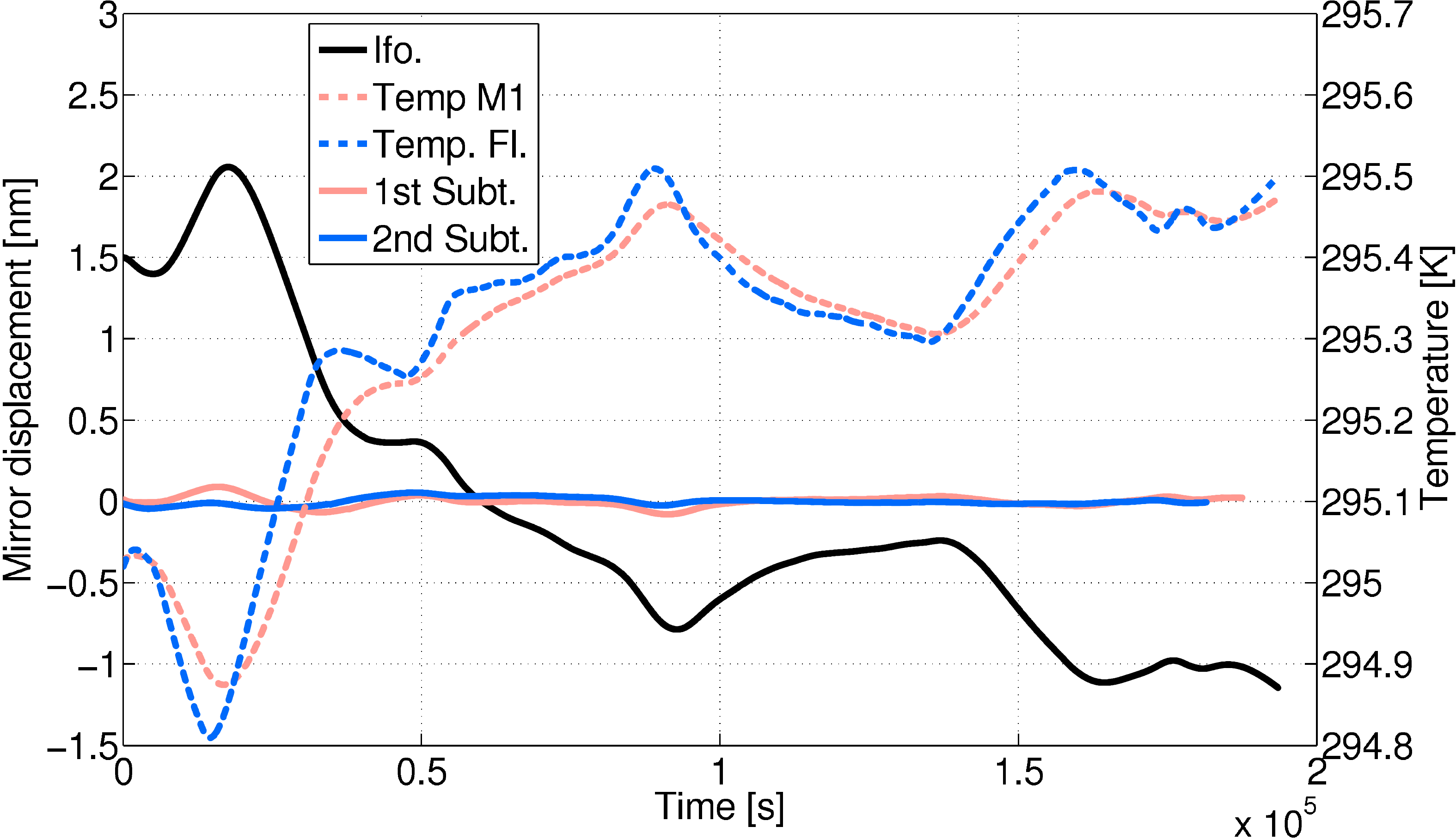}
\includegraphics[width=0.42\textwidth]{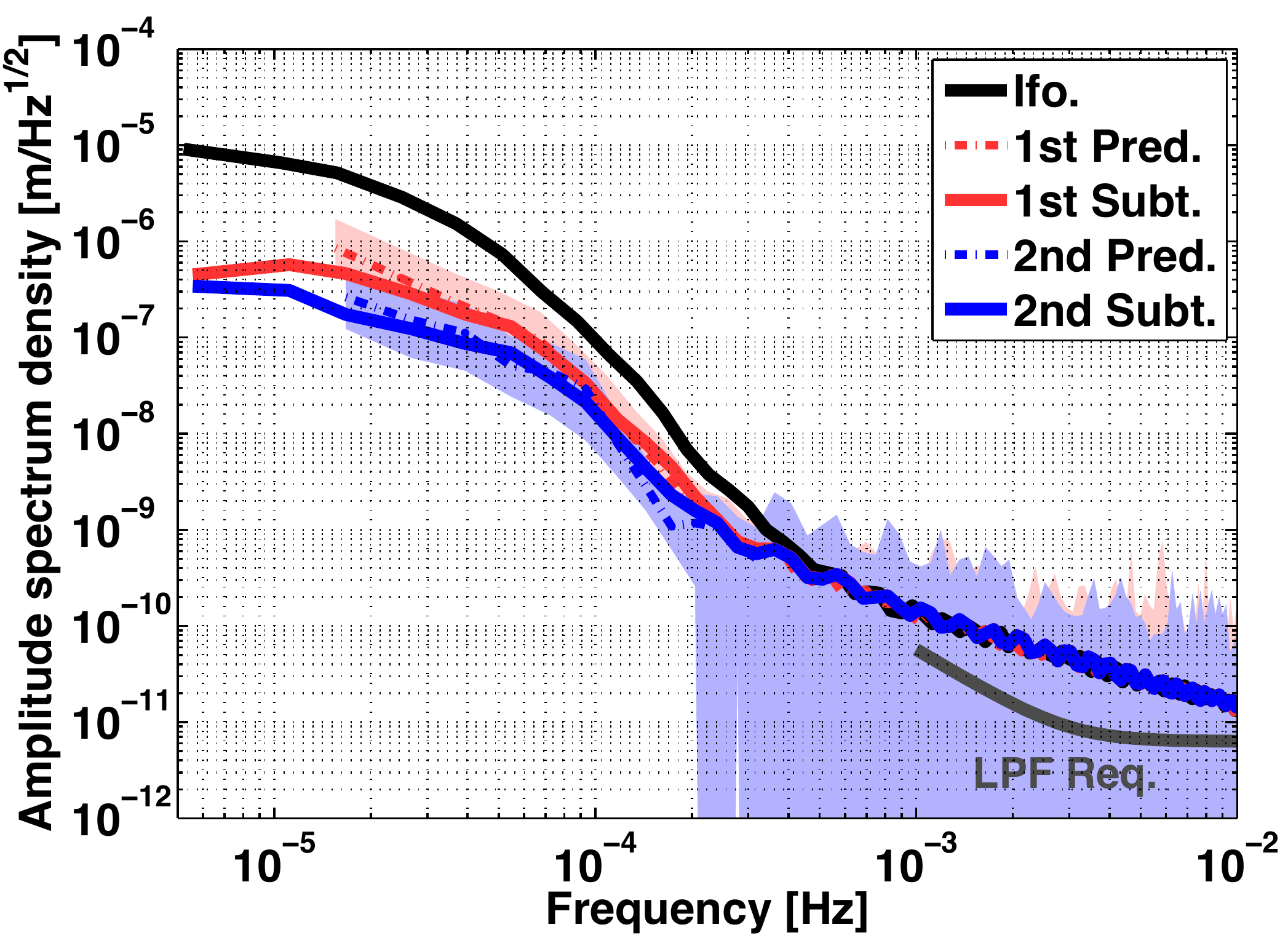}
\caption{Temperature noise subtraction for a two days measurement without
control loops suppressing the interferometric noise contributions. 
{\it Left:} Time domain data series for the two most relevant temperature sensors 
and the interferometer read out. We also show interferometer data after 
subtracting the information contained in the two previous temperatures sensors.
{\it Right:} Comparison of the frequency domain analysis ---previously compared with 
the optimal prediction in Fig.~\ref{fig.cond2opt}--- with the spectra of 
the interferometer time series after the subtraction of the temperature contribution
coming from the two most relevant temperature sensors. For comparison, we also show the
requirement on interferometric noise in the LISA Pathfinder mission.}
\label{fig.subs}
\end{center}
\end{figure*}

A second possible approach is to subtract (in a linear least square sense) 
one by one those contributions perturbing our measurement. This scheme 
leads to simpler expressions, which will 
allow the derivation of the digital filters required to clean the data in the time domain.
This approach, known as \emph{conditioned} spectral analysis, derives 
the dependence between one of the inputs and the output when the other 
inputs are turned off, implicitly requiring here that the correlations between 
inputs are turned off as well. 
In order to introduce this analysis scheme, we refer back to Eq.~(\ref{eq.system})
which we can write now as
\begin{equation}
\Phi(\omega) = \sum^q_{i=1} L_{T_{i}\,\Phi}(\omega)\,T_{i \cdot (i-1)!}(\omega) + n(\omega),
\end{equation} 
where $T_{i \cdot (i-1)!}(\omega)$ represents the Fourier transform of variable 
$T_{i}(t)$ when the linear effects of $T_1$ to $T_{i-1}$ 
have been sequentially removed from $T_i$ by optimum linear least-squares techniques.
Notice that each of this ordered conditioned records will be mutually uncorrelated, a property which
is not generally satisfied by the original records.
The frequency response function $L_{T_{i} \Phi}$ is the optimum linear system to predict 
$\Phi (t)$ from $T_{i} (t)$. Analogously as we did in Eq.(\ref{eq.Snn}), we can derive the optimum 
operator, defined as the one that minimises $S_{nn}$ for any possible combination of $L_{T_{i} \Phi}$. 
This leads to~\cite{Bendat}
\begin{equation}
L_{T_{i} \, \Phi } = \frac{S_{T_i \Phi \cdot T_{(i-1)!}}}{S_{T_i T_i\cdot T_{(i-1)!}}}
\label{eq.Lcond}
\end{equation}
where the spectra and cross-spectra appearing in the previous expressions are
computed on the conditioned variables.
Notice that for the case of a system where q = 1, Eq.~(\ref{eq.Lcond}) and 
Eq.~(\ref{eq.Hoptimal}) will naturally reduce to the same expression
\begin{equation}
L_{T_{1} \, \Phi} = H_1 = \frac{S_{T_1 \Phi}}{S_{T_1  T_1}}
\end{equation}

Following Eq.~(\ref{eq.coherence}) we can define the coherence function for conditioned variables as
\begin{equation}
\gamma^2_{T_i \Phi \cdot T_{(i-1)!}} = \frac{|S_{T_i \Phi \cdot T_{(i-1)!}}|^2}{S_{T_i T_i \cdot T_{(i-1)!}} S_{\Phi \Phi \cdot T_{(i-1)!}}}
\end{equation}
i.e., the coherence between $\Phi$ and $T_i$, 
once all contribution from the previous $i-1$ sensors, $T_{(i-1)!}$, are subtracted.
Together with Eq.~\ref{eq.Lcond}, it can be shown that the conditioned spectra can
be written down in terms of partial coherence functions as follows~\cite{Bendat}
\begin{equation}
 S_{\Phi \Phi \cdot T_{i!}}  =   S_{\Phi \Phi \cdot T_{(i-1)!}} \, \left( 1 - \gamma^2_{ T_i \Phi \cdot T_{(i-1)!} } \right),
\label{eq.cond}
\end{equation} 
where $i=1,2,\ldots q$. This quantity corresponds to the phase noise spectrum, 
$S_{\Phi \Phi}(\omega)$, when contributions from temperature sensors, 
$T_{i}$ are recursively removed by linear least squares.

We will use Eq.~(\ref{eq.cond}) in the following to obtain predictions of the 
expected noise reduction when we subtract a given temperature contribution 
from the main interferometer measurement. This will give us a useful diagnostic
tool to evaluate the contribution coming from each location independently. 
Also, it is worth stressing that, in comparison with the optimal method, the 
sequential approach in the conditioned scheme allows us to deal with expressions
with a maximum of $q$ terms, instead of the $q^2$ linear system required in the 
optimal approach, being $q$ the number of noise contributions in our problem.

\section{Temperature noise subtraction \label{sec.subt}}
\subsection{Frequency domain analysis}
Before removing the contribution arising from temperature, we proceed to estimate
the expected reduction. To do this, we compute the conditioned
spectra.
Results are shown in Fig.
\ref{fig.cond2opt} where the curve labelled as `prediction' shows the
conditioned spectrum in Eq.~(\ref{eq.cond}) 
for the two sensors with the strongest contributions 
---according to Fig.~\ref{fig.tf}. 
We also performed the analysis for a third sensor but the coherence with 
the interferometer was already too small to show any improvement.
In Fig.~\ref{fig.cond2opt} we also compare the conditioned spectra for the two 
first subtractions with the spectrum obtained following the optimal 
subtraction scheme, presented in Sec.~\ref{sec.optimal}.
The comparison confirms that the second subtraction in the conditioned 
scheme achieves the optimal level and, therefore, subsequent 
subtractions will not improve substantially.
As expected, a coherent subtraction
of the temperature contribution would reduce the noise level at frequencies
$\rm f \leq 0.4\,mHz$.  According to this first analysis, the subtraction
of the contribution contained in these two sensors
would reduce the instrument noise floor by a factor 5.6 (15\,dB) at the very low end
of the LISA measurement band, $\rm f = 0.1\,mHz$. The factor increases up to 20.2
(26\,dB) at the lowest frequency bin, $\rm f = 15\,\mu Hz$.

\subsection{Time domain analysis}
We want to be able to subtract noise contributions in time domain, 
which can be useful to avoid effects purely related to the Fourier domain,
e.g. correlation between frequency bins. Time domain analysis 
is of relevance as well in order to obtain temperature noise cleaned
time-series which can be used for subsequent analysis.
For these reasons we have previously
resampled both data streams  with the same sampling frequency and on to a synchronous
time grid. 
As previously shown, the relation between temperature and phase read out is better
described by their frequency domain description.  We then translate this transfer
function into a digital filter, i.e., a recursive relation that allows us
to include a certain dynamical response and, in particular, a delayed action of
the temperature upon the interferometer. The tool used here is the vector fitting
algorithm~\cite{Gustavsen99}  
which allows us to fit the measured transfer function in terms
of N poles, $p_k$, and residues , $z_k$, 
\begin{equation}  
h_{\Phi T_i}(z^{-1}) = \sum_{k=1}^{N} \frac{r_k }{1-p_k\,z^{-1}}.
\label{eq.respol} 
\end{equation}

Since our previous analysis showed that the temperature contribution to the interferometer
is relevant in a frequency range up to $\rm \simeq1\,mHz$, we fit the
transfer function up to this frequency. Figure \ref{fig.tf}  shows the fit result
with a 3rd order model. Once the digital filter is obtained, we filter the temperature
measurement with $h_{\Phi T_i}(z^{-1})$ to obtain the temperature contribution
to the interferometer, which we can then readily subtract from the original
measurement.
This procedure is performed initially considering the temperature reading
of the sensor attached to the mounting of the first mirror ($\rm T_{\rm M1}$). Then, the information
contained in the flange sensor ($\rm T_{\rm FL}$) is removed from the residual of the first subtraction.

\begin{figure}[t]
\includegraphics[width=0.5\textwidth]{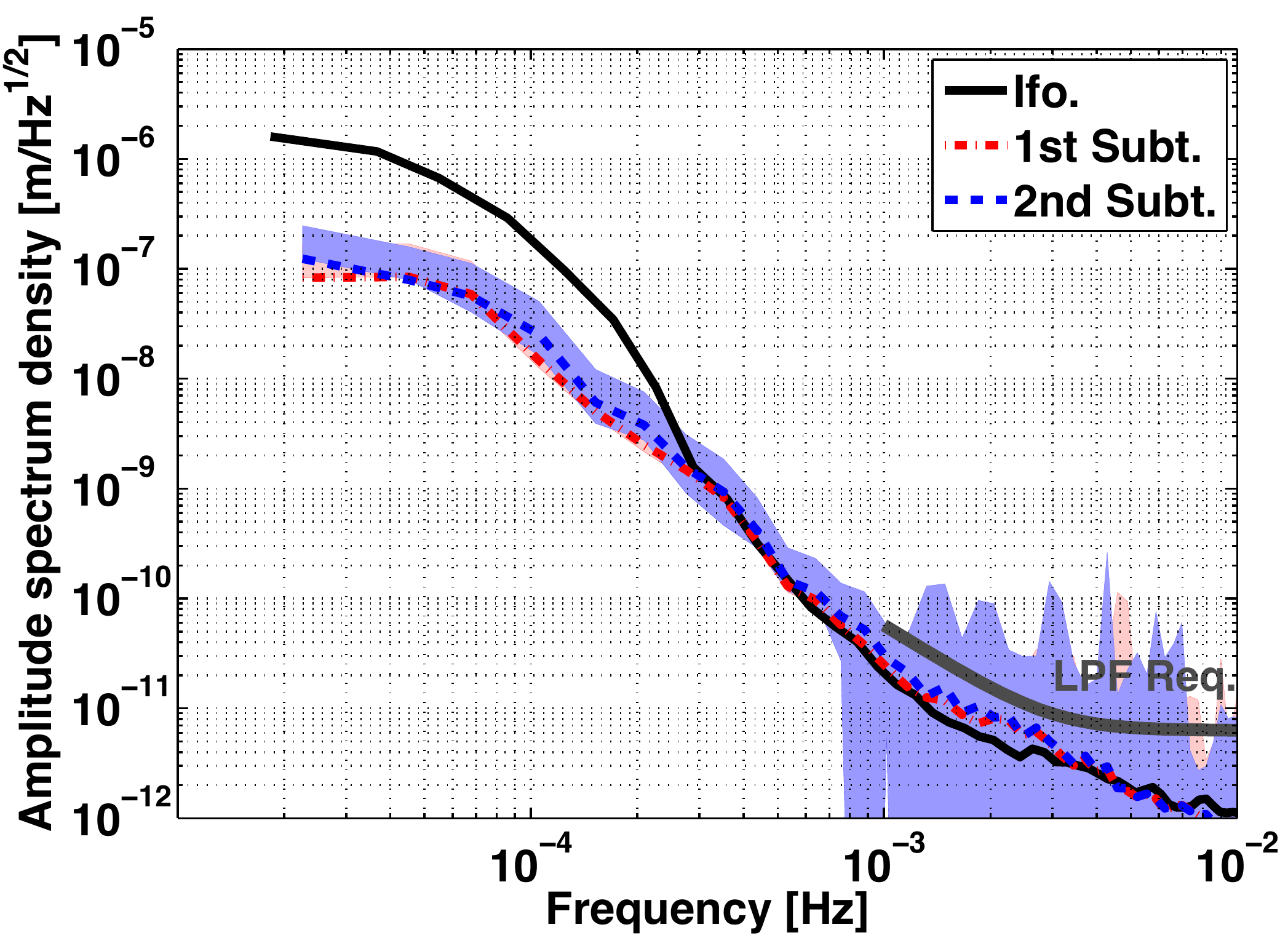}
\caption{Temperature noise substraction for a data segment where
control loops were active. As in previous analysis, 
we show the subtraction for the two more relevant sensors
and we compare with the LISA Pathfinder interferometer  
noise requirement, showing compliance in this run.}
\label{fig.subs_performance}
\end{figure}

As shown in Fig.~\ref{fig.subs}, the subtracted curve is in agreement with the 
one previously obtained in the frequency
domain analysis. We show, for comparison, the LISA Pathfinder interferometric
noise goal on the same plot. Since the control loops were not suppressing the
interferometric noise contributions, the measurement is above the goal. To further
investigate the temperature noise contribution in our setup we used the method
above to subtract the temperature noise contribution in a measurement where the
instrument was operating in closed-loop (all noise suppression on). The transfer
function is the same as the one previously evaluated since the setup remains
unchanged. As shown in Fig.~\ref{fig.subs_performance}, the effect of the subtraction
reduces the noise level at lower frequencies, $\rm f \leq 0.2\,mHz$, when compared 
with the control-free case although
the effect is stronger at $\rm f = 0.1\,mHz$, reducing the noise floor a factor
10 (20\,dB) after the subtraction. We notice as well that, according to
our current analysis, we can not attribute the complete noise contribution
in the low end of the LISA
Pathfinder measuring band, $\rm f \simeq 1\,mHz$, to temperature driven phase fluctuations. \\


%

\section{Discussion and Conclusions \label{sec.concl}}

The analysis reported here characterises, for the first time, the low frequency temperature
coupling to an interferometric measurement in a realistic gravitational
wave instrumental setup. We also introduce a methodology to
clean the main measurement of this noise contribution by removing the
information contained in multiple sensors monitoring the experiment.
Once applied, we obtain a noise-free time series whose spectral 
density confirms the prediction obtained by a frequency domain noise projection. 
We have shown as well that the proposed method, based on conditioned data streams,
agrees with the prediction obtained with an optimal approach. The advantage 
of the former method being a simplification in the required analysis as 
well as a straightforward characterization of the contribution of each temperature 
sensor independently. 
Results on the LISA Pathfinder optical bench show that, when considering the temperature readings
in our current setup, the instrument performance is not affected in the Pathfinder measurement band
after the temperature noise subtraction, although it gives
a significant noise reduction at lower frequencies, and therefore this noise source has a direct
impact for LISA-like experiments. This is of relevance for any ground-based gravitational
 wave experiments aiming at low frequencies, and in particular for those testing
technologies for space-based gravitational wave observatories. 
It is worth stressing that, after temperature noise subtraction, our experiment 
shows a performance of $100\, \rm nm/\sqrt{Hz}$ at f = 30 $\rm  \mu Hz$.
This precision at very low frequencies in an optical metrology measurement can only 
be compared with results from a recent test campaign of the LISA Pathfinder 
spacecraft in an artificial space environment~\cite{Guzman13},
which further confirms that the method described here can allow on-ground lab 
experiments to compute a performance curve without
the limitation of the unavoidable temperature drifts,
achieving results which are close to space conditions. 
The method is extendable as well to any frequency dependent contribution 
that is independently measurable in gravitational wave experiments.

\acknowledgments

We gratefully acknowledge support by 
Deutsches Zentrum f\"ur Luft- und Raumfahrt (DLR) 
(reference~50~OQ~0601).
We also acknowledge support from contracts AYA-2010-15709 
(MICINN) and 2009-SGR-935 (AGAUR). 
NK acknowledges support from the FI PhD program of the
AGAUR (Generalitat de Catalunya). MN acknowledges
a JAE-doc grant from CSIC and support from the EU
Marie Curie CIG 322288.



\bibliographystyle{apsrev4-1}
\bibliography{library}

\begin{thebibliography}{17}%
\makeatletter
\providecommand \@ifxundefined [1]{%
 \@ifx{#1\undefined}
}%
\providecommand \@ifnum [1]{%
 \ifnum #1\expandafter \@firstoftwo
 \else \expandafter \@secondoftwo
 \fi
}%
\providecommand \@ifx [1]{%
 \ifx #1\expandafter \@firstoftwo
 \else \expandafter \@secondoftwo
 \fi
}%
\providecommand \natexlab [1]{#1}%
\providecommand \enquote  [1]{``#1''}%
\providecommand \bibnamefont  [1]{#1}%
\providecommand \bibfnamefont [1]{#1}%
\providecommand \citenamefont [1]{#1}%
\providecommand \href@noop [0]{\@secondoftwo}%
\providecommand \href [0]{\begingroup \@sanitize@url \@href}%
\providecommand \@href[1]{\@@startlink{#1}\@@href}%
\providecommand \@@href[1]{\endgroup#1\@@endlink}%
\providecommand \@sanitize@url [0]{\catcode `\\12\catcode `\$12\catcode
  `\&12\catcode `\#12\catcode `\^12\catcode `\_12\catcode `\%12\relax}%
\providecommand \@@startlink[1]{}%
\providecommand \@@endlink[0]{}%
\providecommand \url  [0]{\begingroup\@sanitize@url \@url }%
\providecommand \@url [1]{\endgroup\@href {#1}{\urlprefix }}%
\providecommand \urlprefix  [0]{URL }%
\providecommand \Eprint [0]{\href }%
\providecommand \doibase [0]{http://dx.doi.org/}%
\providecommand \selectlanguage [0]{\@gobble}%
\providecommand \bibinfo  [0]{\@secondoftwo}%
\providecommand \bibfield  [0]{\@secondoftwo}%
\providecommand \translation [1]{[#1]}%
\providecommand \BibitemOpen [0]{}%
\providecommand \bibitemStop [0]{}%
\providecommand \bibitemNoStop [0]{.\EOS\space}%
\providecommand \EOS [0]{\spacefactor3000\relax}%
\providecommand \BibitemShut  [1]{\csname bibitem#1\endcsname}%
\let\auto@bib@innerbib\@empty
\bibitem [{\citenamefont {Abadie}\ \emph {et~al.}(2010)\citenamefont {Abadie}
  \emph {et~al.}}]{Abadie10}%
  \BibitemOpen
  \bibfield  {author} {\bibinfo {author} {\bibfnamefont {J.}~\bibnamefont
  {Abadie}} \emph {et~al.} (\bibinfo {collaboration} {The LIGO Scientific
  Collaboration and The Virgo Collaboration}),\ }\href {\doibase
  10.1103/PhysRevD.81.102001} {\bibfield  {journal} {\bibinfo  {journal} {Phys.
  Rev. D}\ }\textbf {\bibinfo {volume} {81}},\ \bibinfo {pages} {102001}
  (\bibinfo {year} {2010})}\BibitemShut {NoStop}%
\bibitem [{\citenamefont {Bender}\ \emph {et~al.}(2000)\citenamefont {Bender}
  \emph {et~al.}}]{Bender00}%
  \BibitemOpen
  \bibfield  {author} {\bibinfo {author} {\bibfnamefont {P.}~\bibnamefont
  {Bender}} \emph {et~al.},\ }\href@noop {} {\emph {\bibinfo {title} {Laser
  Interferometer Space Antenna: a cornerstone mission for the observation of
  gravitational waves}}},\ \bibinfo {type} {Tech. Rep.}\ \bibinfo {number}
  {ESA-SCI(2000)11}\ (\bibinfo  {institution} {ESA},\ \bibinfo {year}
  {2000})\BibitemShut {NoStop}%
\bibitem [{\citenamefont {{Amaro-Seoane}}\ \emph {et~al.}(2012)\citenamefont
  {{Amaro-Seoane}}, \citenamefont {{Aoudia}}, \citenamefont {{Babak}},
  \citenamefont {{Bin{\'e}truy}}, \citenamefont {{Berti}}, \citenamefont
  {{Boh{\'e}}}, \citenamefont {{Caprini}}, \citenamefont {{Colpi}},
  \citenamefont {{Cornish}}, \citenamefont {{Danzmann}}, \citenamefont
  {{Dufaux}}, \citenamefont {{Gair}}, \citenamefont {{Jennrich}}, \citenamefont
  {{Jetzer}}, \citenamefont {{Klein}}, \citenamefont {{Lang}}, \citenamefont
  {{Lobo}}, \citenamefont {{Littenberg}}, \citenamefont {{McWilliams}},
  \citenamefont {{Nelemans}}, \citenamefont {{Petiteau}}, \citenamefont
  {{Porter}}, \citenamefont {{Schutz}}, \citenamefont {{Sesana}}, \citenamefont
  {{Stebbins}}, \citenamefont {{Sumner}}, \citenamefont {{Vallisneri}},
  \citenamefont {{Vitale}}, \citenamefont {{Volonteri}},\ and\ \citenamefont
  {{Ward}}}]{Amaro12}%
  \BibitemOpen
  \bibfield  {author} {\bibinfo {author} {\bibfnamefont {P.}~\bibnamefont
  {{Amaro-Seoane}}}, \bibinfo {author} {\bibfnamefont {S.}~\bibnamefont
  {{Aoudia}}}, \bibinfo {author} {\bibfnamefont {S.}~\bibnamefont {{Babak}}},
  \bibinfo {author} {\bibfnamefont {P.}~\bibnamefont {{Bin{\'e}truy}}},
  \bibinfo {author} {\bibfnamefont {E.}~\bibnamefont {{Berti}}}, \bibinfo
  {author} {\bibfnamefont {A.}~\bibnamefont {{Boh{\'e}}}}, \bibinfo {author}
  {\bibfnamefont {C.}~\bibnamefont {{Caprini}}}, \bibinfo {author}
  {\bibfnamefont {M.}~\bibnamefont {{Colpi}}}, \bibinfo {author} {\bibfnamefont
  {N.~J.}\ \bibnamefont {{Cornish}}}, \bibinfo {author} {\bibfnamefont
  {K.}~\bibnamefont {{Danzmann}}}, \bibinfo {author} {\bibfnamefont {J.-F.}\
  \bibnamefont {{Dufaux}}}, \bibinfo {author} {\bibfnamefont {J.}~\bibnamefont
  {{Gair}}}, \bibinfo {author} {\bibfnamefont {O.}~\bibnamefont {{Jennrich}}},
  \bibinfo {author} {\bibfnamefont {P.}~\bibnamefont {{Jetzer}}}, \bibinfo
  {author} {\bibfnamefont {A.}~\bibnamefont {{Klein}}}, \bibinfo {author}
  {\bibfnamefont {R.~N.}\ \bibnamefont {{Lang}}}, \bibinfo {author}
  {\bibfnamefont {A.}~\bibnamefont {{Lobo}}}, \bibinfo {author} {\bibfnamefont
  {T.}~\bibnamefont {{Littenberg}}}, \bibinfo {author} {\bibfnamefont {S.~T.}\
  \bibnamefont {{McWilliams}}}, \bibinfo {author} {\bibfnamefont
  {G.}~\bibnamefont {{Nelemans}}}, \bibinfo {author} {\bibfnamefont
  {A.}~\bibnamefont {{Petiteau}}}, \bibinfo {author} {\bibfnamefont {E.~K.}\
  \bibnamefont {{Porter}}}, \bibinfo {author} {\bibfnamefont {B.~F.}\
  \bibnamefont {{Schutz}}}, \bibinfo {author} {\bibfnamefont {A.}~\bibnamefont
  {{Sesana}}}, \bibinfo {author} {\bibfnamefont {R.}~\bibnamefont
  {{Stebbins}}}, \bibinfo {author} {\bibfnamefont {T.}~\bibnamefont
  {{Sumner}}}, \bibinfo {author} {\bibfnamefont {M.}~\bibnamefont
  {{Vallisneri}}}, \bibinfo {author} {\bibfnamefont {S.}~\bibnamefont
  {{Vitale}}}, \bibinfo {author} {\bibfnamefont {M.}~\bibnamefont
  {{Volonteri}}}, \ and\ \bibinfo {author} {\bibfnamefont {H.}~\bibnamefont
  {{Ward}}},\ }\href@noop {} {\bibfield  {journal} {\bibinfo  {journal} {ArXiv
  e-prints}\ } (\bibinfo {year} {2012})},\ \Eprint
  {http://arxiv.org/abs/1201.3621} {arXiv:1201.3621 [astro-ph.CO]} \BibitemShut
  {NoStop}%
\bibitem [{\citenamefont {Peabody}\ and\ \citenamefont
  {Merkowitz}(2005)}]{Peabody05}%
  \BibitemOpen
  \bibfield  {author} {\bibinfo {author} {\bibfnamefont {H.}~\bibnamefont
  {Peabody}}\ and\ \bibinfo {author} {\bibfnamefont {S.}~\bibnamefont
  {Merkowitz}},\ }\href@noop {} {\bibfield  {journal} {\bibinfo  {journal}
  {Class. Quant. Grav.}\ }\textbf {\bibinfo {volume} {22}},\ \bibinfo {pages}
  {S403} (\bibinfo {year} {2005})}\BibitemShut {NoStop}%
\bibitem [{\citenamefont {Carbone}\ \emph {et~al.}(2007)\citenamefont {Carbone}
  \emph {et~al.}}]{Carbone07}%
  \BibitemOpen
  \bibfield  {author} {\bibinfo {author} {\bibnamefont {Carbone}} \emph
  {et~al.},\ }\href {\doibase 10.1103/PhysRevD.76.102003} {\bibfield  {journal}
  {\bibinfo  {journal} {Phys. Rev. D}\ }\textbf {\bibinfo {volume} {76}},\
  \bibinfo {pages} {102003} (\bibinfo {year} {2007})}\BibitemShut {NoStop}%
\bibitem [{\citenamefont {Cavalleri}\ \emph {et~al.}(2009)\citenamefont
  {Cavalleri} \emph {et~al.}}]{Cavalleri09b}%
  \BibitemOpen
  \bibfield  {author} {\bibinfo {author} {\bibnamefont {Cavalleri}} \emph
  {et~al.},\ }\href {\doibase 10.1103/PhysRevLett.103.140601} {\bibfield
  {journal} {\bibinfo  {journal} {Phys. Rev. Lett.}\ }\textbf {\bibinfo
  {volume} {103}},\ \bibinfo {pages} {140601} (\bibinfo {year}
  {2009})}\BibitemShut {NoStop}%
\bibitem [{\citenamefont {Bender}(2003)}]{Bender03}%
  \BibitemOpen
  \bibfield  {author} {\bibinfo {author} {\bibfnamefont {P.~L.}\ \bibnamefont
  {Bender}},\ }\href@noop {} {\bibfield  {journal} {\bibinfo  {journal} {Class.
  Quant. Grav.}\ }\textbf {\bibinfo {volume} {20}},\ \bibinfo {pages} {S301}
  (\bibinfo {year} {2003})}\BibitemShut {NoStop}%
\bibitem [{\citenamefont {Anza}\ \emph {et~al.}(2005)\citenamefont {Anza} \emph
  {et~al.}}]{Anza05}%
  \BibitemOpen
  \bibfield  {author} {\bibinfo {author} {\bibfnamefont {S.}~\bibnamefont
  {Anza}} \emph {et~al.},\ }\href {\doibase 10.1088/0264-9381/22/10/001}
  {\bibfield  {journal} {\bibinfo  {journal} {Class. Quant. Grav.}\ }\textbf
  {\bibinfo {volume} {22}},\ \bibinfo {pages} {S125} (\bibinfo {year}
  {2005})}\BibitemShut {NoStop}%
\bibitem [{\citenamefont {Armano}\ \emph {et~al.}(2009)\citenamefont {Armano}
  \emph {et~al.}}]{Armano09}%
  \BibitemOpen
  \bibfield  {author} {\bibinfo {author} {\bibfnamefont {M.}~\bibnamefont
  {Armano}} \emph {et~al.},\ }\href {\doibase 10.1088/0264-9381/26/9/094001}
  {\bibfield  {journal} {\bibinfo  {journal} {Class. Quant. Grav.}\ }\textbf
  {\bibinfo {volume} {26}},\ \bibinfo {pages} {094001} (\bibinfo {year}
  {2009})}\BibitemShut {NoStop}%
\bibitem [{\citenamefont {Lobo}\ \emph {et~al.}(2006)\citenamefont {Lobo},
  \citenamefont {Nofrarias}, \citenamefont {Ramos-Castro},\ and\ \citenamefont
  {Sanjuan}}]{Lobo06a}%
  \BibitemOpen
  \bibfield  {author} {\bibinfo {author} {\bibfnamefont {A.}~\bibnamefont
  {Lobo}}, \bibinfo {author} {\bibfnamefont {M.}~\bibnamefont {Nofrarias}},
  \bibinfo {author} {\bibfnamefont {J.}~\bibnamefont {Ramos-Castro}}, \ and\
  \bibinfo {author} {\bibfnamefont {J.}~\bibnamefont {Sanjuan}},\ }\href
  {\doibase 10.1088/0264-9381/23/17/005} {\bibfield  {journal} {\bibinfo
  {journal} {Class. Quant. Grav.}\ }\textbf {\bibinfo {volume} {23}},\ \bibinfo
  {pages} {5177} (\bibinfo {year} {2006})},\ \Eprint
  {http://arxiv.org/abs/gr-qc/0603102} {arXiv:gr-qc/0603102} \BibitemShut
  {NoStop}%
\bibitem [{\citenamefont {Sanju{\'a}n}\ \emph {et~al.}(2009)\citenamefont
  {Sanju{\'a}n}, \citenamefont {Ramos-Castro},\ and\ \citenamefont
  {Lobo}}]{Sanjuan09b}%
  \BibitemOpen
  \bibfield  {author} {\bibinfo {author} {\bibfnamefont {J.}~\bibnamefont
  {Sanju{\'a}n}}, \bibinfo {author} {\bibfnamefont {J.}~\bibnamefont
  {Ramos-Castro}}, \ and\ \bibinfo {author} {\bibfnamefont {A.}~\bibnamefont
  {Lobo}},\ }\href@noop {} {\bibfield  {journal} {\bibinfo  {journal} {Class.
  Quant. Grav.}\ }\textbf {\bibinfo {volume} {26}},\ \bibinfo {pages} {094009}
  (\bibinfo {year} {2009})}\BibitemShut {NoStop}%
\bibitem [{\citenamefont {Heinzel}\ \emph {et~al.}(2003)\citenamefont {Heinzel}
  \emph {et~al.}}]{Heinzel03}%
  \BibitemOpen
  \bibfield  {author} {\bibinfo {author} {\bibfnamefont {G.}~\bibnamefont
  {Heinzel}} \emph {et~al.},\ }\href@noop {} {\bibfield  {journal} {\bibinfo
  {journal} {Class. Quantum Grav.}\ }\textbf {\bibinfo {volume} {20}},\
  \bibinfo {pages} {153} (\bibinfo {year} {2003})}\BibitemShut {NoStop}%
\bibitem [{\citenamefont {Heinzel}\ \emph {et~al.}(2005)\citenamefont {Heinzel}
  \emph {et~al.}}]{Heinzel05}%
  \BibitemOpen
  \bibfield  {author} {\bibinfo {author} {\bibfnamefont {G.}~\bibnamefont
  {Heinzel}} \emph {et~al.},\ }\href@noop {} {\bibfield  {journal} {\bibinfo
  {journal} {Class. Quant. Grav.}\ }\textbf {\bibinfo {volume} {22}},\ \bibinfo
  {pages} {S149} (\bibinfo {year} {2005})}\BibitemShut {NoStop}%
\bibitem [{\citenamefont {Audley}\ \emph {et~al.}(2011)\citenamefont {Audley},
  \citenamefont {Danzmann}, \citenamefont {Mar{\'\i}n}, \citenamefont
  {Heinzel}, \citenamefont {Monsky}, \citenamefont {Nofrarias}, \citenamefont
  {Steier}, \citenamefont {Gerardi}, \citenamefont {Gerndt}, \citenamefont
  {Hechenblaikner}, \citenamefont {Johann}, \citenamefont {Luetzow-Wentzky},
  \citenamefont {Wand}, \citenamefont {Antonucci}, \citenamefont {Armano},
  \citenamefont {Auger}, \citenamefont {Benedetti}, \citenamefont {Binetruy},
  \citenamefont {Boatella}, \citenamefont {Bogenstahl}, \citenamefont
  {Bortoluzzi}, \citenamefont {Bosetti}, \citenamefont {Caleno}, \citenamefont
  {Cavalleri}, \citenamefont {Cesa}, \citenamefont {Chmeissani}, \citenamefont
  {Ciani}, \citenamefont {Conchillo}, \citenamefont {Congedo}, \citenamefont
  {Cristofolini}, \citenamefont {Cruise}, \citenamefont {Marchi}, \citenamefont
  {Diaz-Aguilo}, \citenamefont {Diepholz}, \citenamefont {Dixon}, \citenamefont
  {Dolesi}, \citenamefont {Fauste}, \citenamefont {Ferraioli}, \citenamefont
  {Fertin}, \citenamefont {Fichter}, \citenamefont {Fitzsimons}, \citenamefont
  {Freschi}, \citenamefont {Marirrodriga}, \citenamefont {Gesa}, \citenamefont
  {Gibert}, \citenamefont {Giardini}, \citenamefont {Grimani}, \citenamefont
  {Grynagier}, \citenamefont {Guillaume}, \citenamefont {Guzm{\'a}n},
  \citenamefont {Harrison}, \citenamefont {Hewitson}, \citenamefont
  {Hollington}, \citenamefont {Hough}, \citenamefont {Hoyland}, \citenamefont
  {Hueller}, \citenamefont {Huesler}, \citenamefont {Jeannin}, \citenamefont
  {Jennrich}, \citenamefont {Jetzer}, \citenamefont {Johlander}, \citenamefont
  {Killow}, \citenamefont {Llamas}, \citenamefont {Lloro}, \citenamefont
  {Lobo}, \citenamefont {Maarschalkerweerd}, \citenamefont {Madden},
  \citenamefont {Mance}, \citenamefont {Mateos}, \citenamefont {McNamara},
  \citenamefont {Mendes}, \citenamefont {Mitchell}, \citenamefont {Nicolini},
  \citenamefont {Nicolodi}, \citenamefont {Pedersen}, \citenamefont
  {Perreur-Lloyd}, \citenamefont {Perreca}, \citenamefont {Plagnol},
  \citenamefont {Prat}, \citenamefont {Racca}, \citenamefont {Rais},
  \citenamefont {Ramos-Castro}, \citenamefont {Reiche}, \citenamefont {Perez},
  \citenamefont {Robertson}, \citenamefont {Rozemeijer}, \citenamefont
  {Sanjuan}, \citenamefont {Schulte}, \citenamefont {Shaul}, \citenamefont
  {Stagnaro}, \citenamefont {Strandmoe}, \citenamefont {Sumner}, \citenamefont
  {Taylor}, \citenamefont {Texier}, \citenamefont {Trenkel}, \citenamefont
  {Tombolato}, \citenamefont {Vitale}, \citenamefont {Wanner}, \citenamefont
  {Ward}, \citenamefont {Waschke}, \citenamefont {Wass}, \citenamefont
  {Weber},\ and\ \citenamefont {Zweifel}}]{Audley11}%
  \BibitemOpen
  \bibfield  {author} {\bibinfo {author} {\bibfnamefont {H.}~\bibnamefont
  {Audley}}, \bibinfo {author} {\bibfnamefont {K.}~\bibnamefont {Danzmann}},
  \bibinfo {author} {\bibfnamefont {A.~G.}\ \bibnamefont {Mar{\'\i}n}},
  \bibinfo {author} {\bibfnamefont {G.}~\bibnamefont {Heinzel}}, \bibinfo
  {author} {\bibfnamefont {A.}~\bibnamefont {Monsky}}, \bibinfo {author}
  {\bibfnamefont {M.}~\bibnamefont {Nofrarias}}, \bibinfo {author}
  {\bibfnamefont {F.}~\bibnamefont {Steier}}, \bibinfo {author} {\bibfnamefont
  {D.}~\bibnamefont {Gerardi}}, \bibinfo {author} {\bibfnamefont
  {R.}~\bibnamefont {Gerndt}}, \bibinfo {author} {\bibfnamefont
  {G.}~\bibnamefont {Hechenblaikner}}, \bibinfo {author} {\bibfnamefont
  {U.}~\bibnamefont {Johann}}, \bibinfo {author} {\bibfnamefont
  {P.}~\bibnamefont {Luetzow-Wentzky}}, \bibinfo {author} {\bibfnamefont
  {V.}~\bibnamefont {Wand}}, \bibinfo {author} {\bibfnamefont {F.}~\bibnamefont
  {Antonucci}}, \bibinfo {author} {\bibfnamefont {M.}~\bibnamefont {Armano}},
  \bibinfo {author} {\bibfnamefont {G.}~\bibnamefont {Auger}}, \bibinfo
  {author} {\bibfnamefont {M.}~\bibnamefont {Benedetti}}, \bibinfo {author}
  {\bibfnamefont {P.}~\bibnamefont {Binetruy}}, \bibinfo {author}
  {\bibfnamefont {C.}~\bibnamefont {Boatella}}, \bibinfo {author}
  {\bibfnamefont {J.}~\bibnamefont {Bogenstahl}}, \bibinfo {author}
  {\bibfnamefont {D.}~\bibnamefont {Bortoluzzi}}, \bibinfo {author}
  {\bibfnamefont {P.}~\bibnamefont {Bosetti}}, \bibinfo {author} {\bibfnamefont
  {M.}~\bibnamefont {Caleno}}, \bibinfo {author} {\bibfnamefont
  {A.}~\bibnamefont {Cavalleri}}, \bibinfo {author} {\bibfnamefont
  {M.}~\bibnamefont {Cesa}}, \bibinfo {author} {\bibfnamefont {M.}~\bibnamefont
  {Chmeissani}}, \bibinfo {author} {\bibfnamefont {G.}~\bibnamefont {Ciani}},
  \bibinfo {author} {\bibfnamefont {A.}~\bibnamefont {Conchillo}}, \bibinfo
  {author} {\bibfnamefont {G.}~\bibnamefont {Congedo}}, \bibinfo {author}
  {\bibfnamefont {I.}~\bibnamefont {Cristofolini}}, \bibinfo {author}
  {\bibfnamefont {M.}~\bibnamefont {Cruise}}, \bibinfo {author} {\bibfnamefont
  {F.~D.}\ \bibnamefont {Marchi}}, \bibinfo {author} {\bibfnamefont
  {M.}~\bibnamefont {Diaz-Aguilo}}, \bibinfo {author} {\bibfnamefont
  {I.}~\bibnamefont {Diepholz}}, \bibinfo {author} {\bibfnamefont
  {G.}~\bibnamefont {Dixon}}, \bibinfo {author} {\bibfnamefont
  {R.}~\bibnamefont {Dolesi}}, \bibinfo {author} {\bibfnamefont
  {J.}~\bibnamefont {Fauste}}, \bibinfo {author} {\bibfnamefont
  {L.}~\bibnamefont {Ferraioli}}, \bibinfo {author} {\bibfnamefont
  {D.}~\bibnamefont {Fertin}}, \bibinfo {author} {\bibfnamefont
  {W.}~\bibnamefont {Fichter}}, \bibinfo {author} {\bibfnamefont
  {E.}~\bibnamefont {Fitzsimons}}, \bibinfo {author} {\bibfnamefont
  {M.}~\bibnamefont {Freschi}}, \bibinfo {author} {\bibfnamefont {C.~G.}\
  \bibnamefont {Marirrodriga}}, \bibinfo {author} {\bibfnamefont
  {L.}~\bibnamefont {Gesa}}, \bibinfo {author} {\bibfnamefont {F.}~\bibnamefont
  {Gibert}}, \bibinfo {author} {\bibfnamefont {D.}~\bibnamefont {Giardini}},
  \bibinfo {author} {\bibfnamefont {C.}~\bibnamefont {Grimani}}, \bibinfo
  {author} {\bibfnamefont {A.}~\bibnamefont {Grynagier}}, \bibinfo {author}
  {\bibfnamefont {B.}~\bibnamefont {Guillaume}}, \bibinfo {author}
  {\bibfnamefont {F.}~\bibnamefont {Guzm{\'a}n}}, \bibinfo {author}
  {\bibfnamefont {I.}~\bibnamefont {Harrison}}, \bibinfo {author}
  {\bibfnamefont {M.}~\bibnamefont {Hewitson}}, \bibinfo {author}
  {\bibfnamefont {D.}~\bibnamefont {Hollington}}, \bibinfo {author}
  {\bibfnamefont {J.}~\bibnamefont {Hough}}, \bibinfo {author} {\bibfnamefont
  {D.}~\bibnamefont {Hoyland}}, \bibinfo {author} {\bibfnamefont
  {M.}~\bibnamefont {Hueller}}, \bibinfo {author} {\bibfnamefont
  {J.}~\bibnamefont {Huesler}}, \bibinfo {author} {\bibfnamefont
  {O.}~\bibnamefont {Jeannin}}, \bibinfo {author} {\bibfnamefont
  {O.}~\bibnamefont {Jennrich}}, \bibinfo {author} {\bibfnamefont
  {P.}~\bibnamefont {Jetzer}}, \bibinfo {author} {\bibfnamefont
  {B.}~\bibnamefont {Johlander}}, \bibinfo {author} {\bibfnamefont
  {C.}~\bibnamefont {Killow}}, \bibinfo {author} {\bibfnamefont
  {X.}~\bibnamefont {Llamas}}, \bibinfo {author} {\bibfnamefont
  {I.}~\bibnamefont {Lloro}}, \bibinfo {author} {\bibfnamefont
  {A.}~\bibnamefont {Lobo}}, \bibinfo {author} {\bibfnamefont {R.}~\bibnamefont
  {Maarschalkerweerd}}, \bibinfo {author} {\bibfnamefont {S.}~\bibnamefont
  {Madden}}, \bibinfo {author} {\bibfnamefont {D.}~\bibnamefont {Mance}},
  \bibinfo {author} {\bibfnamefont {I.}~\bibnamefont {Mateos}}, \bibinfo
  {author} {\bibfnamefont {P.~W.}\ \bibnamefont {McNamara}}, \bibinfo {author}
  {\bibfnamefont {J.}~\bibnamefont {Mendes}}, \bibinfo {author} {\bibfnamefont
  {E.}~\bibnamefont {Mitchell}}, \bibinfo {author} {\bibfnamefont
  {D.}~\bibnamefont {Nicolini}}, \bibinfo {author} {\bibfnamefont
  {D.}~\bibnamefont {Nicolodi}}, \bibinfo {author} {\bibfnamefont
  {F.}~\bibnamefont {Pedersen}}, \bibinfo {author} {\bibfnamefont
  {M.}~\bibnamefont {Perreur-Lloyd}}, \bibinfo {author} {\bibfnamefont
  {A.}~\bibnamefont {Perreca}}, \bibinfo {author} {\bibfnamefont
  {E.}~\bibnamefont {Plagnol}}, \bibinfo {author} {\bibfnamefont
  {P.}~\bibnamefont {Prat}}, \bibinfo {author} {\bibfnamefont {G.~D.}\
  \bibnamefont {Racca}}, \bibinfo {author} {\bibfnamefont {B.}~\bibnamefont
  {Rais}}, \bibinfo {author} {\bibfnamefont {J.}~\bibnamefont {Ramos-Castro}},
  \bibinfo {author} {\bibfnamefont {J.}~\bibnamefont {Reiche}}, \bibinfo
  {author} {\bibfnamefont {J.~A.~R.}\ \bibnamefont {Perez}}, \bibinfo {author}
  {\bibfnamefont {D.}~\bibnamefont {Robertson}}, \bibinfo {author}
  {\bibfnamefont {H.}~\bibnamefont {Rozemeijer}}, \bibinfo {author}
  {\bibfnamefont {J.}~\bibnamefont {Sanjuan}}, \bibinfo {author} {\bibfnamefont
  {M.}~\bibnamefont {Schulte}}, \bibinfo {author} {\bibfnamefont
  {D.}~\bibnamefont {Shaul}}, \bibinfo {author} {\bibfnamefont
  {L.}~\bibnamefont {Stagnaro}}, \bibinfo {author} {\bibfnamefont
  {S.}~\bibnamefont {Strandmoe}}, \bibinfo {author} {\bibfnamefont {T.~J.}\
  \bibnamefont {Sumner}}, \bibinfo {author} {\bibfnamefont {A.}~\bibnamefont
  {Taylor}}, \bibinfo {author} {\bibfnamefont {D.}~\bibnamefont {Texier}},
  \bibinfo {author} {\bibfnamefont {C.}~\bibnamefont {Trenkel}}, \bibinfo
  {author} {\bibfnamefont {D.}~\bibnamefont {Tombolato}}, \bibinfo {author}
  {\bibfnamefont {S.}~\bibnamefont {Vitale}}, \bibinfo {author} {\bibfnamefont
  {G.}~\bibnamefont {Wanner}}, \bibinfo {author} {\bibfnamefont
  {H.}~\bibnamefont {Ward}}, \bibinfo {author} {\bibfnamefont {S.}~\bibnamefont
  {Waschke}}, \bibinfo {author} {\bibfnamefont {P.}~\bibnamefont {Wass}},
  \bibinfo {author} {\bibfnamefont {W.~J.}\ \bibnamefont {Weber}}, \ and\
  \bibinfo {author} {\bibfnamefont {P.}~\bibnamefont {Zweifel}},\ }\href
  {http://stacks.iop.org/0264-9381/28/i=9/a=094003} {\bibfield  {journal}
  {\bibinfo  {journal} {Class. Quant. Grav.}\ }\textbf {\bibinfo {volume}
  {28}},\ \bibinfo {pages} {094003} (\bibinfo {year} {2011})}\BibitemShut
  {NoStop}%
\bibitem [{\citenamefont {Bendat}\ and\ \citenamefont
  {Piersol}(1993)}]{Bendat}%
  \BibitemOpen
  \bibfield  {author} {\bibinfo {author} {\bibfnamefont {J.~S.}\ \bibnamefont
  {Bendat}}\ and\ \bibinfo {author} {\bibfnamefont {A.~G.}\ \bibnamefont
  {Piersol}},\ }\href@noop {} {\emph {\bibinfo {title} {Engineering
  Applications of Correlation and Spectral Analysis}}}\ (\bibinfo  {publisher}
  {John Wiley \& Sons, Inc.},\ \bibinfo {year} {1993})\BibitemShut {NoStop}%
\bibitem [{\citenamefont {Gustavsen}\ and\ \citenamefont
  {Semlyen}(1999)}]{Gustavsen99}%
  \BibitemOpen
  \bibfield  {author} {\bibinfo {author} {\bibfnamefont {B.}~\bibnamefont
  {Gustavsen}}\ and\ \bibinfo {author} {\bibfnamefont {A.}~\bibnamefont
  {Semlyen}},\ }\href@noop {} {\bibfield  {journal} {\bibinfo  {journal} {IEEE
  Trans. Power Delivery}\ }\textbf {\bibinfo {volume} {14}},\ \bibinfo {pages}
  {1052} (\bibinfo {year} {1999})}\BibitemShut {NoStop}%
\bibitem [{\citenamefont {Cervantes}\ \emph {et~al.}(2013)\citenamefont
  {Cervantes}, \citenamefont {Flatscher}, \citenamefont {Gerardi},
  \citenamefont {Burkhardt}, \citenamefont {Gerndt}, \citenamefont {Nofrarias},
  \citenamefont {Reiche}, \citenamefont {Heinzel}, \citenamefont {Danzmann},
  \citenamefont {Bot{\'e}} \emph {et~al.}}]{Guzman13}%
  \BibitemOpen
  \bibfield  {author} {\bibinfo {author} {\bibfnamefont {F.}~\bibnamefont
  {Cervantes}}, \bibinfo {author} {\bibfnamefont {R.}~\bibnamefont
  {Flatscher}}, \bibinfo {author} {\bibfnamefont {D.}~\bibnamefont {Gerardi}},
  \bibinfo {author} {\bibfnamefont {J.}~\bibnamefont {Burkhardt}}, \bibinfo
  {author} {\bibfnamefont {R.}~\bibnamefont {Gerndt}}, \bibinfo {author}
  {\bibfnamefont {M.}~\bibnamefont {Nofrarias}}, \bibinfo {author}
  {\bibfnamefont {J.}~\bibnamefont {Reiche}}, \bibinfo {author} {\bibfnamefont
  {G.}~\bibnamefont {Heinzel}}, \bibinfo {author} {\bibfnamefont
  {K.}~\bibnamefont {Danzmann}}, \bibinfo {author} {\bibfnamefont
  {L.}~\bibnamefont {Bot{\'e}}},  \emph {et~al.},\ }in\ \href@noop {} {\emph
  {\bibinfo {booktitle} {Astronomical Society of the Pacific Conference
  Series}}},\ Vol.\ \bibinfo {volume} {467}\ (\bibinfo {year} {2013})\ p.\
  \bibinfo {pages} {141}\BibitemShut {NoStop}%
\end{thebibliography}%

\end{document}